\documentclass[useAMS,usenatbib]{mn2e}

\usepackage{graphicx}
\usepackage{times}
\usepackage{natbib}
\usepackage{color}

\newcommand{\apjs}{ApJS}

\newcommand{\apjl}{ApJL}

\newcommand{\aap}{A \& A}

\newcommand{\enzo}{\it{\small ENZO}}

\begin{document}
 
\title[Simulated dynamo in the ICM] {Resolved magnetic dynamo action in the simulated intracluster medium}
\author[F. Vazza, G. Brunetti, M. Br\"{u}ggen and A. Bonafede]{F. Vazza$^{1,2,3}$\thanks{E-mail: franco.vazza2@unibo.it}, G. Brunetti$^{1}$, M. Br\"{u}ggen$^{2}$ and A. Bonafede$^{1,2}$\\
$^{1}$Istituto di Radio Astronomia, INAF, Via Godetti 101, 40121 Bologna, Italy\\
$^{2}$ Hamburger Sternwarte, Gojenbergsweg 112, 21029 Hamburg, Germany\\
$^{3}$ Dipartimento di Fisica e Astronomia, Universit\'{a} di Bologna, Via Govetti 92/3, 40121, Bologna, Italy}
\date{Received / Accepted}
\maketitle
\begin{abstract}

Faraday rotation and synchrotron emission from extragalactic radio sources give evidence for the presence of magnetic fields extending over $\sim$ Mpc scales. However, the origin of these fields remains elusive. 
With new high-resolution grid simulations we studied the  growth of magnetic fields in a massive galaxy cluster that in several aspects is similar to the Coma cluster. We investigated models in which magnetic fields originate from primordial seed fields with comoving strengths of $0.1 ~\rm nG$ at redshift $z=30$. The simulations show evidence of significant magnetic field amplification. At the best spatial resolution ($3.95 ~\rm kpc$), we are able to resolve the scale where magnetic tension balances the bending of magnetic lines by turbulence. This allows us to observe the final growth stage of the small-scale dynamo. To our knowledge this is  the first time that this is seen in cosmological simulations  of the intracluster medium. Our mock observations of Faraday Rotation provide a good match to observations of the Coma cluster. 
However, the distribution of magnetic fields shows strong departures from a simple Maxwellian distribution, suggesting that the three-dimensional structure of magnetic fields in real clusters may be significantly different than what is usually assumed when inferring magnetic field values from rotation measure observations. 

\end{abstract}

\label{firstpage} 

\section{Introduction}
\label{sec:intro}

The shape, strength and structure of magnetic fields in galaxy clusters have been inferred from radio observations \citep[e.g.][]{2001ApJ...547L.111C,ct02,fe12}, indicating the presence of $\sim \rm Mpc$-wide fields with a strength of a few $\sim \mu G$.

The distribution of magnetic fields in the intracluster medium (ICM) can be probed via rotation measure (RM) of polarised sources emitting through the cluster volume. For sources located in the background of the cluster, the signal goes as $\rm RM \propto  \int B_{\ ||} n_e dl$, where $B_{\ ||}$ is the magnetic field component along the line of sight, $n_e$ is the electron density and $dl$ is the line element along the line of sight. From the observed distribution of sources across a cluster it is possible to 
infer the three-dimensional distribution of magnetic fields that reproduces the data best \citep[e.g.][]{2001A&A...379..807G,2001ApJ...547L.111C,mu04,gu08,bo10,vacca10,2016A&A...596A..22B}. Central to this method is the assumption that the underlying magnetic fields are isotropic, with Gaussian-distributed components, and that the magnetic field spectrum is 
described by a power law \citep[e.g.][]{1991MNRAS.250..726T,mu04}.
The Coma cluster presently gives the most stringent constraints on the ICM magnetic field, owing to the large number (12) of polarised sources detected in and behind the cluster \citep[][]{bo10,bo13}. The best-fit model for the central $\sim$ Mpc region gives $B(r)=B_0 (n/n_0)^{0.5}$, with $B_0=4.7 \mu G$ ($n_0$ is the core gas density). Observed RM data are consistent with the assumption of a Kolmogorov spectrum of magnetic fluctuations in the $\sim 1-50 ~\rm kpc$ scale range  \citep[][]{bo10}. 
 However, this model underestimates the observed Faraday rotation (FR) signal from the SW sector of Coma, in the direction of the radio relic (Coma C) \citep[][]{bo13}.

The origin of magnetic fields in clusters is not fully understood. High-order fluctuations and non-Gaussianities in the CMB constrain primordial magnetic fields
to be $\leq 10^{-9}$ G (comoving) on scales of $\sim ~\rm Mpc$ at the time of decoupling \citep[][]{PLANCK2015}. Conversely, lower limits on the magnetisation of voids of  $\geq 10^{-16} \rm G$ have been derived from the spectra of high-z blazar sources (e.g.\citealt{2010Sci...328...73N}, see also \citealt{2017arXiv170202586T} for a recent review). The amplification of primordial weak magnetic fields \citep[][]{sub15} via a turbulent dynamo during structure formation might explain the observed magnetic fields inside clusters \citep[e.g.][]{do99,br05,ry08}. Still, other astrophysical sources of magnetic seeding at lower redshift, such as supernovae and active galactic nuclei (AGN), may well contribute to the cosmological magnetic fields seen today \citep[e.g.][]{2006MNRAS.370..319B,donn09, xu09,2017arXiv170601890S}. 
Distinguishing between competing magnetogenesis scenarios in the ICM is almost impossible because the memory of seed fields is believed to be erased by the dynamo \citep[][]{cho14,2016ApJ...817..127B}.   However, these models predict discrepant results at cluster outskirts, in filaments and voids \citep[e.g.][and discussion therein]{donn09,va15radio}. 
With a  large set of new cosmological simulations, we recently explored how competing seeding scenarios for extragalactic magnetic fields affect a wide range of observables and concluded that radio observations offer the best way to constrain magnetogenesis \citep{va17cqg}.
In particular, the Square Kilometre Array (SKA) and its pathfinders (e.g. LOFAR, MWA, MeerKAT and ASKAP), both, in continuum \citep[][]{2011JApA...32..577B,va15ska,va16radio} and polarised \citep[][]{2013A&A...554A.102G,2015arXiv150102298T,2015aska.confE.105G} intensity will have the potential to probe cluster outskirts and the magnetisation of the cosmic web outside of clusters \citep[][]{2015aska.confE..95B}, also using stacking techniques \citep[][]{2015arXiv150100390S} or cross-correlation with galaxy catalogs \citep[][]{vern17,brown17}. 
\\

\begin{figure}
\begin{center}
\includegraphics[width = 0.49\textwidth]{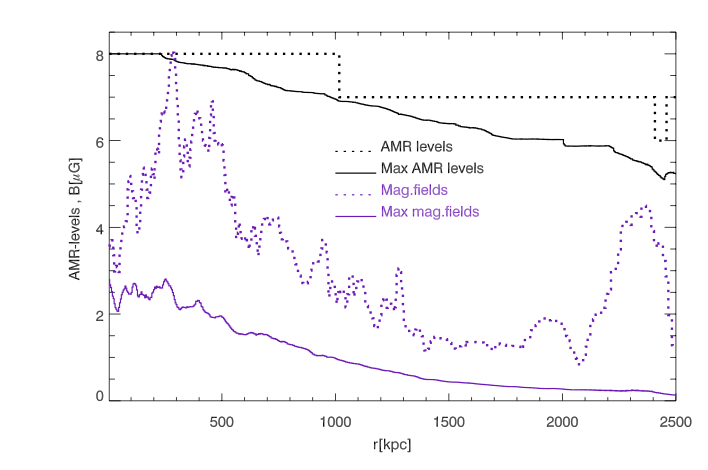}
\caption{Radial distribution of the refinement levels in our AMR8 run at $z=0$ (black lines) and of the magnetic field at the same radius (purple). The solid lines show the average quantities and the dotted lines the maximum values at each radius.}
 \label{fig:AMR_prof}
\end{center}
\end{figure}

\begin{figure*}
\begin{center}
\includegraphics[width = 0.95\textwidth]{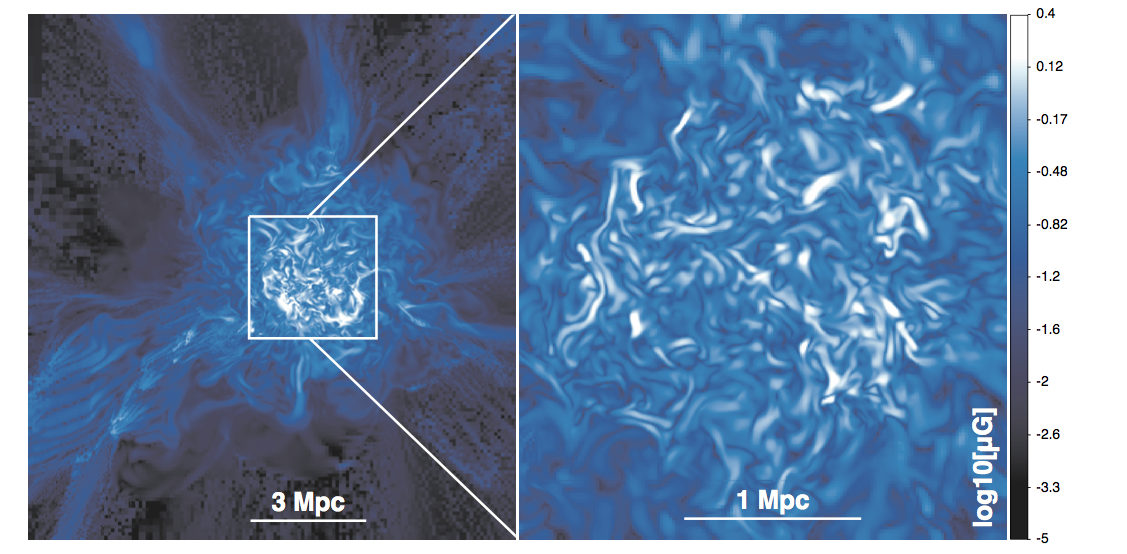}
\caption{Map of projected mean magnetic field for our most resolved run, showing the entire zoom AMR region (left) and a closeup view at the highest resolution (right) for our AMR8 run at $z=0$.}
 \label{map_best}
\end{center}
\end{figure*}

\begin{figure*}
\begin{center}
\includegraphics[width = 0.95\textwidth]{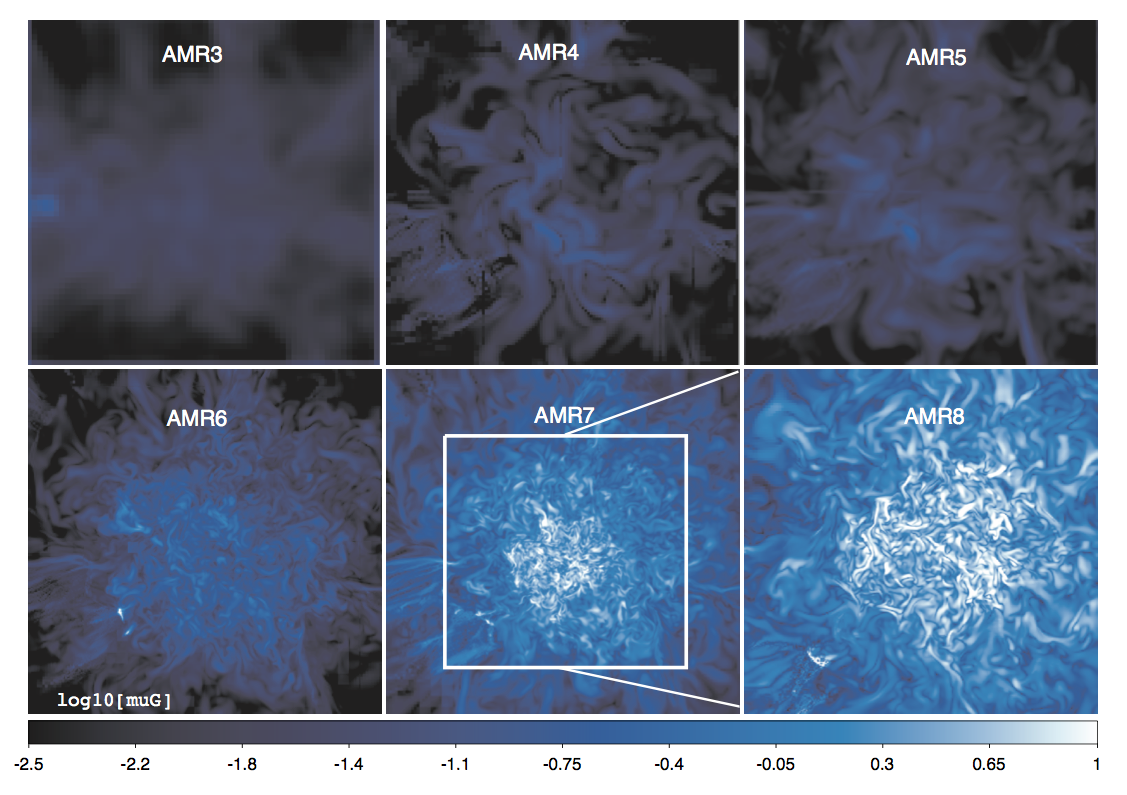}
\caption{Map of projected mean magnetic field strength for resimulations of our cluster at an increasing resolution, for regions of $8.1 \times 8.1 ~\rm Mpc^2$ around the cluster centre at $z=0$.  Each panel shows the mass-weighted magnetic field strength (in units of $\log_{\rm 10} [\mu G]$ for a slice of $\approx 250$ kpc along the line of sight.}
 \label{map_resolution}
\end{center}
\end{figure*}

The direct numerical simulation of dynamo amplification requires high spatial resolution. This is necessary to resolve the scale, $l_A$, below which the magnetic tension can counteract the bending of magnetic field lines and start the linear growth stage of the small-scale dynamo in the intracluster medium  \citep[e.g.][]{2004ApJ...612..276S}. This scale varies with time and across space, as the Universe expands and the magnetic fields and the turbulent motions evolve \citep[e.g.][]{2016ApJ...817..127B}.
 
The growth of the small-scale dynamo  in the intracluster medium has hardly been observed in cosmological grid simulations \citep[e.g.][]{br05,2008A&A...482L..13D,co11,ruszkowski11,va14mhd,2015MNRAS.453.3999M,2016arXiv160105083E,wi17}, owing to the difficulty of reaching a large enough Reynolds number and thus resolve $l_A$. Magnetic field configurations close to the observed ones were only obtained by adding cooling and a local amplification from feedback-induced turbulence \citep[][]{2008A&A...482L..13D,co11,ruszkowski11,2015MNRAS.453.3999M,ma17}, or by explicitly including the additional magnetisation from active galactic nuclei \citep[][]{xu09,2011ApJ...739...77X,2012ApJ...759...40X,sk13,va14mhd}. 

The results from smoothed-particle hydrodynamical (SPH) cosmological simulation disagree with the aforementioned results from grid simulations, in that they typically find a much larger amplification of magnetic fields starting from high redshifts \citep[][]{do99,2007MNRAS.375..657G,do08,donn09,2009MNRAS.398.1678D,bo11,beck12,2013MNRAS.428...13S,beck13}. 

Here, for the first time (as far as we know) we could a) show evidence for the non-linear stage of dynamo amplification in a Eulerian simulation of the ICM {\footnote{We notice that evidence of small-scale dynamo in cosmological simulations of a Milky Way-like galaxy has been recently reported by \citet[][]{2017MNRAS.469.3185P}, using a moving mesh hydrodynamical method.}}; b) produce a good match to the observed FR in the Coma cluster, starting from a primordial weak field. In order to reach the linear dynamo regime, it is crucial that the effective spatial resolution allows the presence of flows with a high hydrodynamical Reynolds number ($R_e \geq 1000$) and to resolve the local MHD scale (Sec.~3.3). We present a number of diagnostics, such as the evolution of distribution functions, power spectra of the magnetic field, as well as the statistics of field curvature, as described in Sec.~3.1-3.3.  
The structure of the paper is the following: in Sec.~2 we present in detail the numerical setup used to simulate magnetic fields in galaxy clusters; in Sec.~3 we show our results, and in Sec.~4 we discuss their possible limitations and relevance for future radio observations. Finally, our conclusions are given in Sec.~5. 

The assumed cosmology in this work a $\Lambda$CDM model with: $h = 0.72$, $\Omega_{\mathrm{M}} = 0.258$, $\Omega_{\mathrm{b}}=0.0441$ and  $\Omega_{\Lambda} = 0.742$, as in \citet[][]{va10kp}.\\

\begin{table}
\label{tab:tab}
\caption{Main parameters of our MHD runs for the cluster simulated in this work. The first column gives the maximum number of AMR levels for each run, the second the corresponding maximum spatial resolution and the third the mass resolution of dark matter particles; the fourth column gives the initial seed field (in comoving units, starting from $z=30$), and the last column gives the ID used throughout the paper}.
\centering \tabcolsep 5pt 
\begin{tabular}{c|c|c|c|c}
  $N_{\rm AMR}$ & $\Delta x_{\rm max}$[kpc] & $m_{\rm DM} \rm [M_{\odot}]$ & seeding & ID\\  \hline 
  $3$ & $126.4$ & $1.04 \cdot 10^{11}$ &  $B_0=0.1$ nG  & AMR3\\
  $4$ & $63.2$ & $1.04 \cdot 10^{11}$ &$B_0=0.1$ nG  & AMR4\\
  $5$ & $31.6$ & $1.04 \cdot 10^{11}$ &$B_0=0.1$ nG  & AMR5\\
  $6$ & $15.8$ & $1.04 \cdot 10^{11}$ &$B_0=0.1$ nG  & AMR6\\
  $7$ & $7.9$ & $1.04 \cdot 10^{11}$ &$B_0=0.1$ nG  & AMR7\\
  $8$ & $3.95$ & $1.04 \cdot 10^{11}$&$B_0=0.1$ nG & AMR8\\
  $8$ & $3.95$ & $1.30 \cdot 10^{10}$ &$B_0=0.1$ nG & AMR8dm\\
$8$ & $3.95$ & $1.30 \cdot 10^{10}$ &$B_{0X}=0.1$ nG & AMR8bx\\ 
$8$ & $3.95$ & $1.30 \cdot 10^{10}$ &$B_0=0.03$ nG & AMR8b0.03nG\\ \hline
  $8$ & $3.95$ & $1.30 \cdot 10^{10}$ &$B_0=0.1$ nG & AMR8dm\_E14\\

  \end{tabular}
\end{table}

\begin{figure}
\begin{center}
\includegraphics[width = 0.495\textwidth]{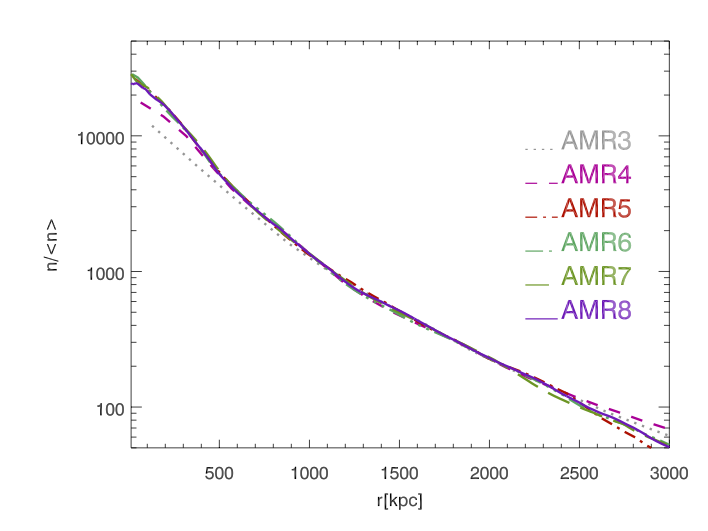}
\includegraphics[width = 0.495\textwidth]{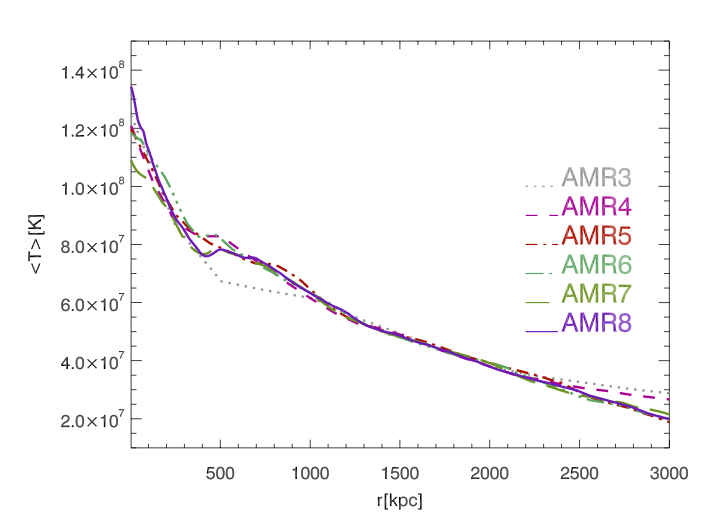}
\includegraphics[width = 0.495\textwidth]{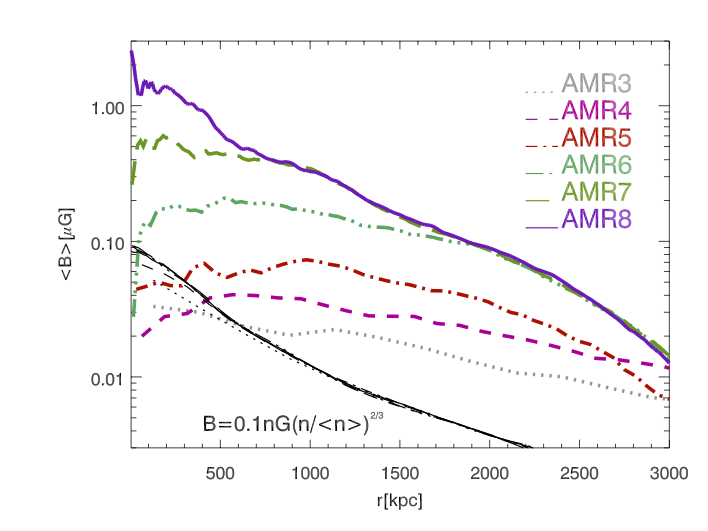}
\caption{Radial profile of gas density (top panel), gas temperature (central panel),
  and magnetic fields (bottom panel) for our reference cluster as a function of resolution at $z=0$. In the third panel, the additional thin black lines show the expected fields in case of simple adiabatic compression of magnetic field lines for every model.}
 \label{prof_res}
\end{center}
\end{figure}

\section{Simulations}
\label{sec:methods}

We simulated the formation of a massive, $\sim 10^{15} M_{\odot}$ galaxy cluster 
using a customised version of the cosmological grid code {\enzo} \citep[][]{enzo13}.  As in our previous work \citep[][]{va14mhd}, we  used the Dedner formulation of MHD equations \citep[][]{wa09} and used adaptive mesh refinement (AMR) to  increase the dynamical resolution \citep[e.g.][]{xu09,2016arXiv160105083E}.

In this work, we only present the result of {\it non-radiative} cosmological simulations that include only the effects of cosmic expansion, gravity and (magneto)hydrodynamics. In forthcoming work, we will simulate intracluster magnetic fields with increasing physical complexity. In this first step, we focus on the role of numerical resolution in the simulation of turbulence and magnetic fields, starting from the simplest magnetic field seed. The limitations of this  model are discussed in Sec.~\ref{numerics}.

This cluster forms in a volume of (260 Mpc)$^3$ (comoving), and is simulated starting from an initially uniform grid of $256^3$ cells and using $256^3$ dark matter particles. The initial density perturbation field is taken from a suite of existing cluster simulations \citep[e.g.][and other works derived from this]{va10kp}. We focused on resimulating this specific object because it has a total mass very close to the mass of the Coma cluster. The Coma cluster is an ideal testbed for magnetic field studies since its large angular extent on the sky allowed for the largest number of RMs from background sources.

The innermost  $\sim$ 25 Mpc$^3$ volume, centred on where the cluster forms, has been further refined using AMR. Refinement is initiated wherever the gas density is $\geq 1\%$ higher than its surroundings. In a resolution study, we produced six resimulations of the same object by increasing the maximum refinement level, to monitor how magnetic field amplification evolves with resolution, from the coarsest resolution of $\Delta x_{\rm max}=126.4 ~\rm kpc$ to the highest resolution of $\Delta x_{\rm max}=3.95 ~\rm kpc$ (see Tab.~1 for a list of of our runs).

Fig.~\ref{fig:AMR_prof} shows the radial profile of the mean and maximum number of AMR levels for the same snapshot (in the same plot, we additionally show also the profiles of mean and maximum magnetic field strength for the cluster). Basically, the entire volume of the cluster is refined at least up to the 6th AMR level ($15.8$ kpc) at $z=0$, and the vast majority of the central volume within $\leq 1$ Mpc from the cluster centre is simulated at the highest possible resolution of 3.95 kpc/cell). 

Unlike previous work \citep[][]{va10kp}, in most of our runs we do not employ nested initial conditions to selectively increase the mass resolution of dark matter particles within the cluster as this would increase the necessary computational resources. Therefore, our mass resolution is limited to $m_{\rm DM}=1.04 \cdot 10^{11} M_{\odot}$. We comment on the effect of a limited mass resolution in Sec.~\ref{sec:discussion}, where we include the comparison with a resimulation of the same object at $8$ times higher mass resolution for the DM component (run AMR8dm in Tab.~1).

All our runs assume a simple primordial seeding scenario, in which we initialised the magnetic field to a uniform value $B_0$ across the entire computational domain, along each coordinate axis. The initial magnetic seed field of $0.1 ~\rm nG$ (comoving) is chosen to be below the upper limits from the analysis of the CMB \citep[e.g.][]{sub15}.
{\footnote{Upper limits of similar strength have been derived for present-day magnetic fields in the intergalactic medium, using statistical analysis of Faraday Rotation at high redshift \citep[e.g.][]{2016PhRvL.116s1302P}. }}. In this work, we simply imposed  the same initial value for each magnetic field component at the starting redshift ($z=30$). This initialisation is obviously simplistic and neglects other possible initial distributions
of magnetic fields, which would be allowed by CMB observations \citep[][]{PLANCK2015}. However, this initialisation can be compared most easily to previous work. As a sanity check, we also tested a few simple variations at the highest resolution, by a) imposing the initial seed field only along the x-axis (keeping the same $B_0$ strength, run AMR8bx in Tab.~1); b) testing a halved initial seed field  ($B_0=0.03 ~\rm nG$, run AMR8b0.03nG). We analyze the outcome of these models in Sec.~\ref{sec:discussion}.

Fig.~\ref{map_best} shows the magnetic field in a thin slice (thickness ($\approx 100 ~\rm kpc$) across the entire (25 Mpc)$^3$ volume where we used AMR, and a close-up view of the innermost 4 Mpc (in this case limited to a thinner slice of $\approx 8 ~\rm kpc$). 

\begin{figure}
\begin{center}
\includegraphics[width = 0.495\textwidth]{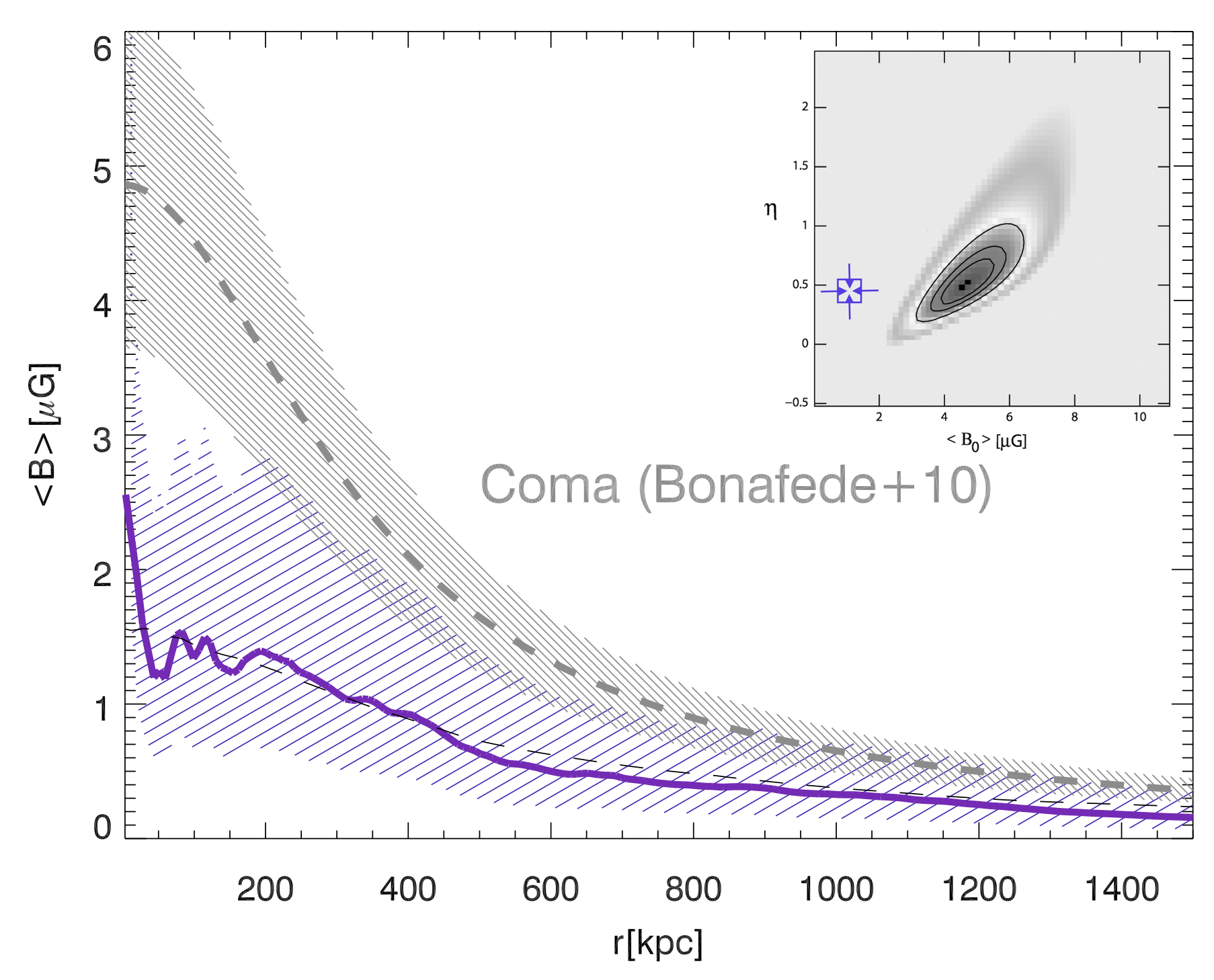}
\caption{Radial profile of magnetic field in our most resolved run AMR8, compared to the magnetic field profile inferred by \citet{bo10} from observed RM. The dashed region of the profiles show the $\pm 1 \sigma$ dispersion around the mean. The dashed line shows the best fit to our measured profile, while the inset show the uncertainty region of the best-fit in \citet{bo10}. The best fit solution for our the profile of our simulated cluster is shown by the purple cross.}
 \label{profB}
\end{center}
\end{figure}

\begin{figure*}
\begin{center}
\includegraphics[width = 0.33\textwidth]{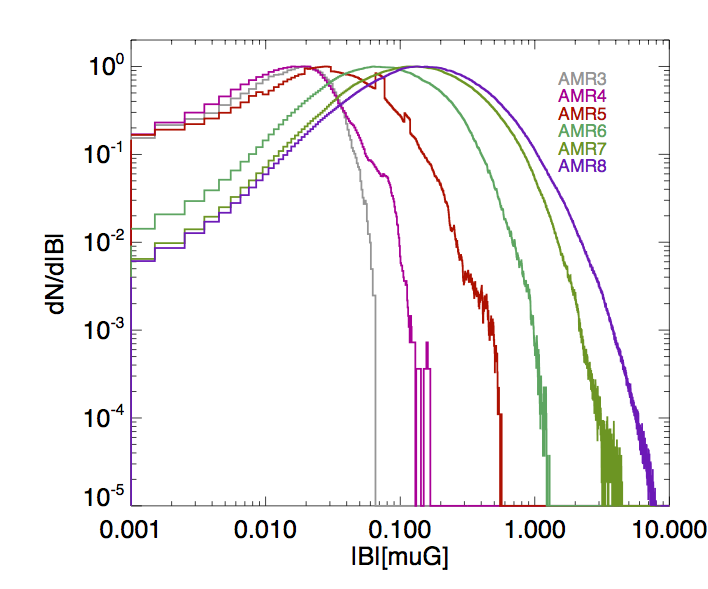}
\includegraphics[width = 0.33\textwidth]{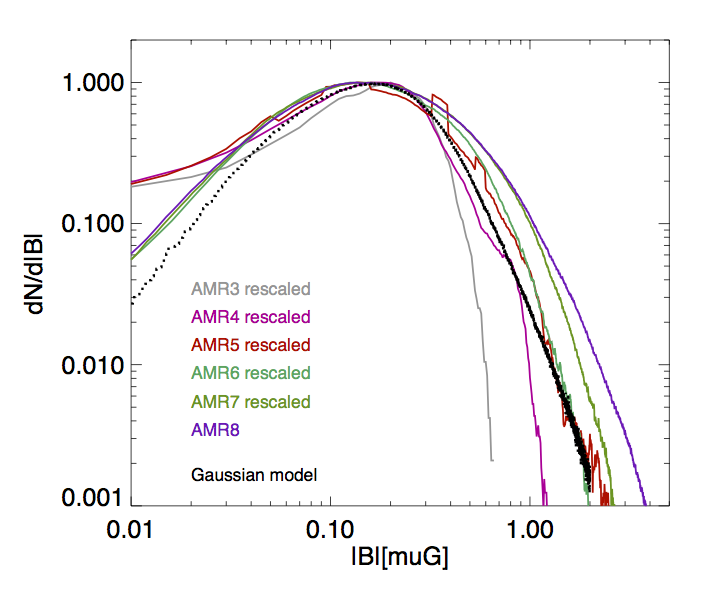}
\includegraphics[width = 0.33\textwidth]{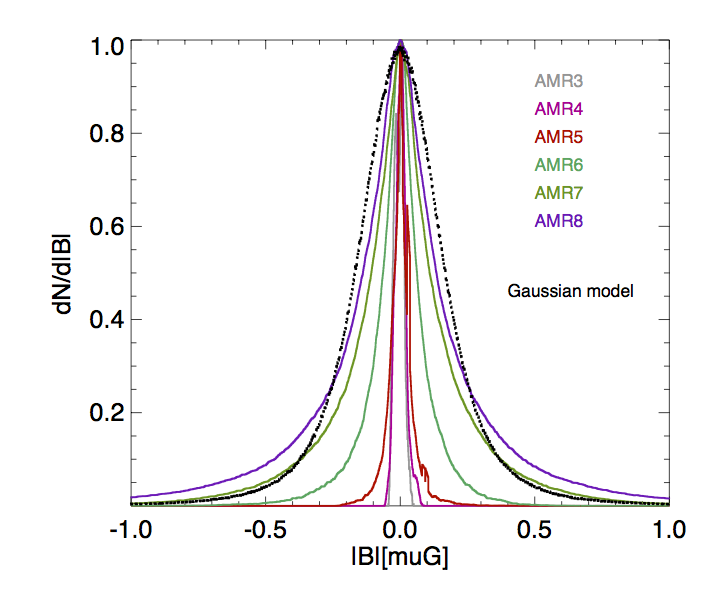}
\caption{Volumetric distribution of magnetic fields in our galaxy cluster, as a function of resolution. Left panel: distribution functions of the magnetic field strength for the central $(1.25 ~\rm Mpc)^3$ volume in our runs at $z=0$.  Central panel:  the same distribution functions of the previous panel, shifted  to match the peak of the AMR8 run, in order to compare the evolution of their shape.  The additional dotted  black line shows the distribution function of magnetic field values for the Gaussian model that best matches the observed profile of Faraday Rotation in the Coma cluster \citep[][]{bo10}, also shifted to match the same peak of the AMR8 run. Right panel: distribution of magnetic field components for all runs; the additional dotted line show the distribution of components for the same Gaussian model of the central panel.}
\label{pdf_res}
\end{center}
\end{figure*}

\begin{figure}
\begin{center}
\includegraphics[width = 0.495\textwidth]{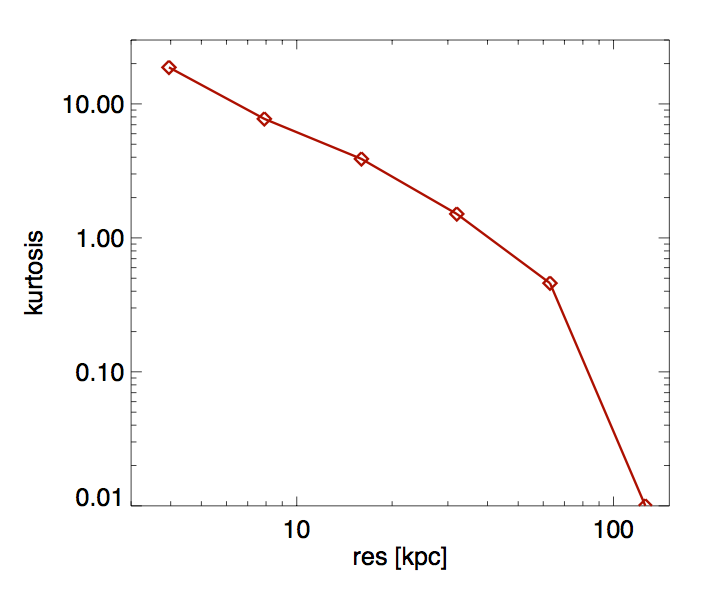}
\caption{Analysis of the departures from Gaussianity in the magnetic field distribution of our runs: mean kurtosis in the distribution of magnetic field components of Fig.~\ref{pdf_res} as a function of spatial resolution.}
\label{kurt_res}
\end{center}
\end{figure}

\begin{figure}
\begin{center}
\includegraphics[width = 0.495\textwidth]{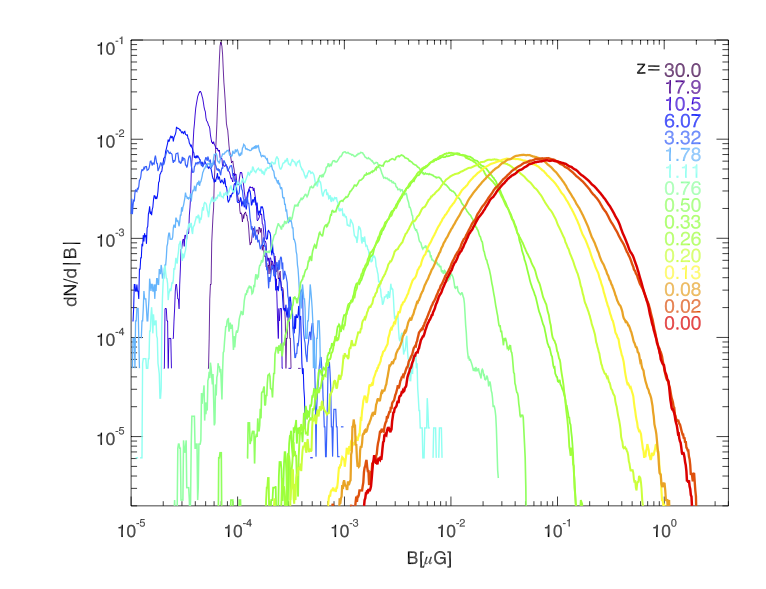}
\caption{Evolution of magnetic fields from $z=30$ to $z=0$ in our galaxy cluster: distribution functions of (comoving) magnetic field strength for the central $\rm Mpc^3$ in our AMR8 run.}
\label{pdf_ev}

\end{center}
\end{figure}
\begin{figure}
\begin{center}
\includegraphics[width = 0.495\textwidth]{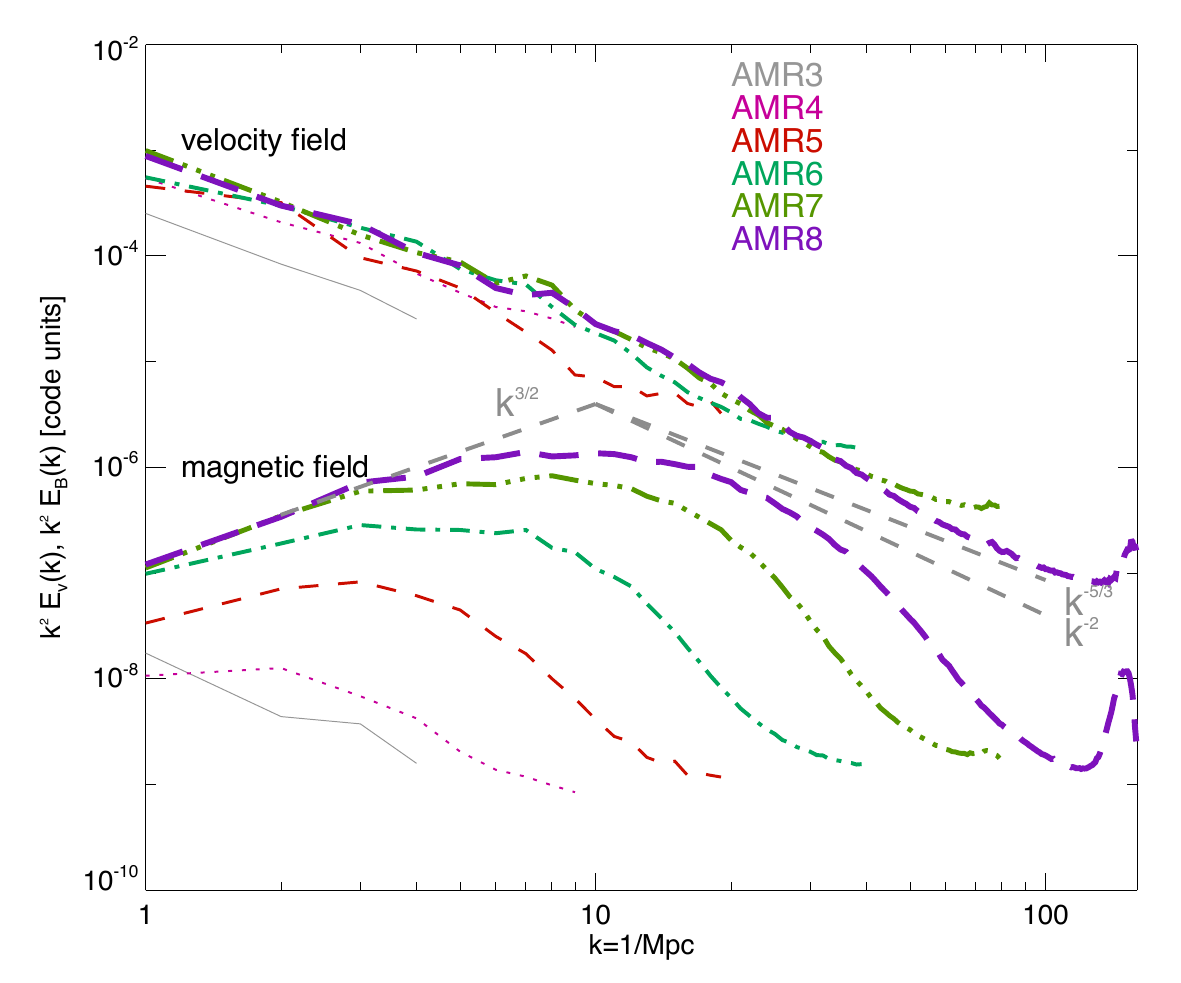}
\caption{Spectral properties of the velocity and magnetic fields of our runs, as a function of resolution. The top lines show the velocity field power spectra (including also the weighting for $n^{0.5}$ as explained in the text), while the bottom lines show the magnetic field power spectra for the innermost (1 Mpc)$^3$ volume of our cluster at $z=0$, for different spatial resolutions. The additional gray lines show the
  $k^{3/2}$ and the $k^{-5/3,-2}$ trends to guide the eye.}
 \label{spec}
\end{center}
\end{figure}

\begin{figure}
\begin{center}
\includegraphics[width = 0.495\textwidth]{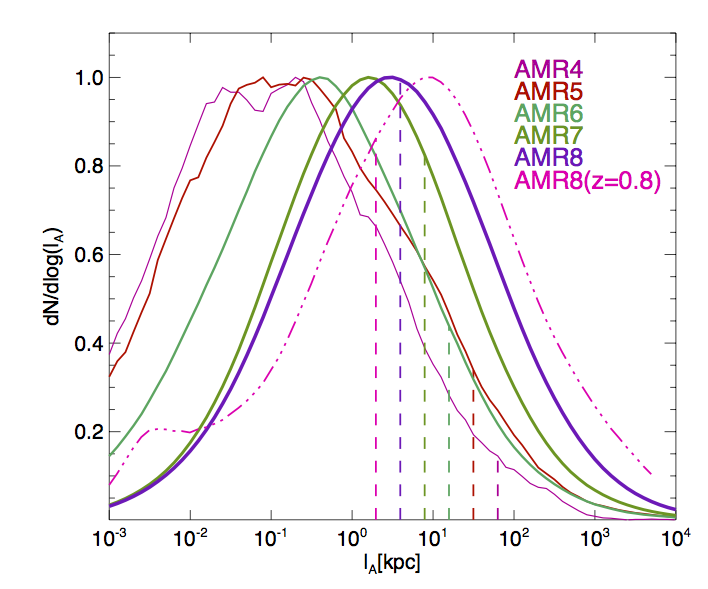}
\caption{Capabilities of our runs in resolving the MHD scale (Eq.~2, see the main text for details). Each histogram gives the distribution of MHD scales in the innermost Mpc$^3$ region in our runs at $z=0$, and additionally for $z=0.8$ in the AMR8 run (dot-dashed line).} 
 \label{scales}
\end{center}
\end{figure}

\section{Results}
\label{sec:res}

\subsection{The volumetric distribution of magnetic fields}
\label{3d}

The map of gas mass-weighted magnetic fields averaged along the line-of-sight for runs with different resolutions and starting from  a reference seed field of $0.1 ~\rm nG$ (comoving, starting from $z=30)$ is shown in Fig.~\ref{map_resolution}, for a line-of-sight of 250 kpc. As is clear from the image, the increase in spatial resolution causes an enhanced level of tangling of the magnetic field lines in the cluster centre, as well as an increased overall magnetic field strength in the innermost volume. Assessing whether this magnetic field growth is indeed a manifestation of the emerging small-scale dynamo amplification in high Reynolds number flow is one of the main goals of this work. 

For each run, we compute the radial profile of the average magnetic field, shown in Fig.~\ref{prof_res} as well as the radial profiles of gas density and gas temperature for reference. While the thermal structure of the cluster appears converged in runs AMR4-AMR8, the magnetic field keeps growing with increasing spatial resolution. 
The mean magnetic field is increased by a factor $\sim 10$ in most volume going from run AMR3 (maximum resolution of 126 kpc) to AMR8 (3.95 kpc). In the cluster core ($\leq 300 ~\rm kpc$) the increase is nearly two orders of magnitude going from AMR3 to AMR8, meaning that the magnetic energy there has increased by $\sim 10^4$ just as a result of the increased resolution.  Since the gas density profile is basically the same in all runs, the excess magnetic field we observe at all radii for increasing resolutions suggests that a dynamo develops as the local Reynolds number of the flow is allowed to be larger ($R_e \propto N^{4/3}$, where $N$ is the 1-dimensional number of cells in the volume, see Sec.\ref{dynamo} for a discussion). This is confirmed by the additional thin black lines in our third panel, which show the expected magnetic field profile in case of simple compression of magnetic field lines ($B=B_0 (n/\langle n \rangle)^{2/3}$, where the density is the one corresponding to each run). The profile of our lowest resolution run is similar to the expectation from pure compression (albeit with a broader distribution, likely as an effect of numerical diffusion) at high resolution the average mean magnetic field is $\geq 30-50$ times larger at all radii compared to the frozen-in prediction. 

In Fig.~\ref{profB}, we plot the magnetic field profile and compare it to the observed profile in the Coma cluster \cite{bo10}, considering also the dispersion around the mean.
The profile of the \citet{bo10} model is directly computed from the 3D model generated  following the recipe in \cite{mu04} and  \citet{bo10}, i.e.  we use the {\it MiR\'{o}} code \citep[][]{bo13} to generate  three-dimensional magnetic field components drawn from a Rayleigh distribution of the vector potential (yielding by construction a Gaussian PDF of magnetic field components and to a Maxwellian distribution of magnetic field strengths),  with a fixed range of spatial wavenumbers and  with a fixed power-law distribution  of magnetic fields (here $P_B \propto k^{-5/3}$). 
At all radii, the dispersion in our simulation is significantly larger than in \citet{bo10}, while the mean field is smaller. However, as we will see in Sec.~\ref{RM}, the simulated RM profile in our AMR8 run gives a good match to the observed RM profile in Coma.  The two profiles only match at the $\sim 2 \sigma$ level around their mean.
Indeed we find that  the radial profile of the AMR8 run can be fitted ($\chi^2=0.00024$) by a simple relation of the kind
\begin{equation}
B(n) \approx 1.55 \rm \mu G \cdot (n/n_0)^{\eta},
\end{equation}
with $n_0=3.5 \cdot 10^{-3}$ cm$^{-3}$ , with $\eta=0.487$, consistent with the model by \citet{bo10} (who find $\eta=1/2$), and a $\sim 3$ times lower normalisation for the magnetic field in the core ($B_0=1.55 ~\rm \mu G$  instead  of $B_0=4.77 \rm ~\mu G$ in \citealt{bo10}).  By comparing with the range of model parameters investigated in \citet{bo10}, as shown in the inset of Fig.~\ref{profB}, we can see that while the slope agrees with the range of values derived from RM observations, the central field value is  entirely outside the region of values constrained by these observations. One of the key results is that the intracluster magnetic fields does not follow a Maxwellian distribution, that would result from a Gaussian distribution of magnetic field components as assumed in the interpretation of Faraday RM \citep[e.g.][]{1991MNRAS.250..726T,mu04,bo10,bo13}.

This is clear in Fig.~\ref{pdf_res} where we show the volume probability distribution function (PDF) of of magnetic field strength  in our runs, extracted from the central $(1.25 ~\rm Mpc)^3$ at $z=0$.

In  the same figure,  we also show the same PDFs, shifted so that their peaks are coincident and the differences in shapes are best highlighted. Finally, the right panel shows the distribution of the magnetic field  components for the same run, in comparison with a purely Gaussian distribution matching the same peak  of the PDF of magnetic field strength of the AMR8 run (obtained as above). The magnetic fields measured in {\it all} our simulations are not Maxwellian/Gaussian, with departures from this model that increase with increasing resolution. In the most resolved  AMR8 run, we observed a tail of values $\sim 2-3$ times larger than in the Maxwellian case, as is highlighted by the central panel. The distribution of magnetic field components also shows the presence of a non-Gaussian tail of values with both signs in the AMR8 run.  

As way to quantify the departure from the Gaussian expectation, we measured the kurtosis of the PDF in our runs (averaged between the various magnetic field components) in (Fig.~\ref{kurt_res}). The values of kurtosis sharply increase with resolution, from a quasi-Gaussian distribution of fields in the AMR3 run to the pronounced non-Gaussianity of the AMR8 run. 

Unlike simpler "turbulence-in-a-box" simulations \citep[e.g.][]{2013MNRAS.429.2469B,2014ApJ...781...84S}, the ICM is an open system where gas with different dynamical histories is continuously mixed. In particular, each merger drives turbulent motions on different scales and with different strengths. In a complex multi-component turbulent flow, the same mechanism also mixes magnetic field components which have been subject to different amplification patterns over their life. Most of the vorticity in the ICM is injected by overdense substructures, which inject vorticity across a wide range of scales and via multiple mergers across several Gyrs \citep[e.g.][]{2017arXiv170602315W}.
The co-existence of different components to the total ICM fields remains visible for $\sim ~\rm Gyr$, reflecting the patchiness of turbulent motions in the ICM \citep[][]{mi14,sch16,va17turbo}.  We notice that these effects are independent of other mechanisms that introduce non-Maxwellian behaviour in turbulent flows, such as intermittency \citep[e.g.][]{2017ApJ...839L..16S}.  As we will discuss in Sec.~\ref{numerics}, the exact shape of the PDF of magnetic field strengths depends quite strongly on the dynamical state of the host cluster. Hence, we do not attempt to provide a fitting formula for it.

\subsection{Evidence for small-scale dynamo}
\label{dynamo}

In order to quantify the effects of the small-scale dynamo, we performed several tests to monitor the temporal, spatial and spectral evolution of the magnetic fields. 

For our most resolved run (AMR8), the evolution of the PDF of magnetic fields in the innermost cluster region across redshifts is shown in Fig.~\ref{pdf_ev}. The evolution is displayed in comoving units{\footnote{$B_{\rm phys} = B_{\rm comoving}/a^2$, where $a$ is the scale factor of the Universe. Displaying the evolution in comoving  units is interesting because it immediately shows the efficiency of the dynamo, which is measured in the comoving reference frame.}}.
The maximum of the PDF steadily grows in time from $\sim 0.1$ nG to $0.1 ~\mu$G. At the same time the PDF broadens until it  develops a non-Maxwellian tail, also characterized by transient features following the accretion of substructures.\\

It is  interesting to investigate whether the spectral signature for a small-scale dynamo amplification is reflected in the power spectra of the magnetic fields for the same runs.
For the same cubic selections of Fig.~\ref{pdf_res} we computed the power spectra (Fig.~\ref{spec}) the velocity field and of the magnetic fields with a standard Fast-Fourier Transform (FFT) approach assuming periodicity (for a discussion see, e.g. \citealt{va17turbo}). In order to allow for a consistent comparison of the kinetic and magnetic energies per mode, the velocity spectra are obtained by multiplying the velocity with $\sqrt{n}$ (where $n$ is the gas density), so that $P_B(k)$ and $P_v(k)$ have the same code units. 
The slope of the power spectra for low wavenumbers is compatible with the Kasantsev model of dynamo $P_B \propto k^{3/2}$ \citep[e.g.][]{2004ApJ...612..276S} while after the peak the spectrum rapidly steepens from $\propto k^{-5/3}$ to $\propto k^{-2}$ or less, consistent with \citep[e.g.][]{2015ApJ...810...93P,2017arXiv170405845R}.
While the shape of the velocity power spectra is hardly modified by a change in resolution, the magnetic field spectra show an abrupt change at the AMR5 run, followed by the formation of a small-scale ($\leq 50~\rm kpc$) pile-up of magnetic energy for runs with a larger resolution. In the AMR8 run, the energy ratio between magnetic and kinetic energy in the $k \sim 10-50$ range (corresponding to $100-20 ~\rm kpc$) reaches a maximum of $\beta_v \sim 0.2-0.3$, i.e. not far from energy equipartition at these modes. The final configuration observed in the AMR8 run confirms that by $z \sim 0$ the small-scale dynamo process has reached the non-linear growth regime, i.e. the magnetic tension at the peak scale is large enough to oppose the further bending of magnetic lines by kinetic pressure on smaller scales. These are also the scales which are mostly responsible for the observed Faraday Rotation (see Sec.\ref{RM}), and are consistent with other recent numerical studies \citep[e.g.][]{2016ApJ...817..127B}.\\

The initial length scale at which magnetic tension can withstand the further bending by hydrodynamic forces can be estimated from the Kolmogorov model of velocity fluctuations as observed in our simulations (Fig.~\ref{spec}) is (Eq. 3 in \citealt{bl07}):

\begin{equation}
l_A \approx 0.3 {\rm kpc} (\frac{B}{\rm \mu G})^3 \frac{L}{\rm kpc} (\frac{n}{10^{-3} \rm part/cm^3})^{-3/2} (\frac{\sigma_L}{\rm km/s})^3 ,
\label{lA}
\end{equation}
where $L$ is a typical eddy size (ideally the injection scale of turbulence) and  $\sigma_L$ is the rms velocity within the scale $L$. Based on this equation, we extracted the turbulent rms velocity within $L=100$ kpc, by filtering out motions on larger scales (with a similar procedure as in \citealt{va17turbo}). Based on this, we computed the 
distribution of $l_A$ for each cell in our central Mpc$^3$ volume at $z=0$, finding the distribution of values given in Fig.~\ref{scales} for various simulations. The vertical dashed lines show the corresponding maximum resolution for each run; only cells to the right of these lines are  resolving the local MHD scale estimated in Eq.~(\ref{lA}).  As resolution is increased, the  critical length  $l_A$ gets resolved in a larger fraction of the volume, up to   $\sim 50\%$  in the AMR8 run. Assessing how this fraction would increase at even larger resolution is difficult. Still, this confirms that at least in a half of our innermost volume, the small-scale dynamo has approached its  linear growth regime. Within the same figure, we show the distribution of $l_A$ limited to run AMR8 at the earlier epoch of $z=0.8$. Interestingly, $l_A$ is resolved for an even larger fraction of the innermost volume at this early epoch. Based on Eq.~2, this is a result of the combined effect of having enhanced turbulence levels below $100 ~\rm kpc$ compared to $z=0$ (due to fast accretions within the forming cluster) as well as the stronger $\propto(1+z)^2$ magnetic fields. We note that, especially prior to the cluster virialization, it is non-trivial to disentangle laminar from turbulent motions, and therefore in this regime the estimate of $l_A$ from Eq.~2 may be overestimated in our approach. In any case, this test suggests that even at high redshift the MHD scale may be resolved already in a significant fraction of the innermost cluster volume, and that therefore the dynamo growth can start soon after the cluster forms. 

Due to the $l_A \propto B^3$ dependence and based on this data, we can also expect that runs with an even slightly increased initial seed field and using the same AMR scheme may enter into the linear growth stage earlier, and for a larger fraction of their volume. However, the analysis of the CMB place the upper limit on the seed fields at the $\sim \rm nG$ level \citep[][]{PLANCK2015}, i.e. only a factor $\sim 10$ above the initial seed field used here. The detailed investigation of how the timing and efficiency of dynamo amplification changes with the seed field values (as well as for other possible seeding mechanisms) will be subject of follow-up work.\\

In the final stage of small-scale dynamo growth, we expect that the curvature of magnetic field lines inversely correlates with their intensity because stronger fields get increasingly harder to bend.
We therefore computed the line-curvature of the
magnetic field distribution following \citet[][]{2004ApJ...612..276S}, $K$, defined as

\begin{equation}
  \vec{K} = \frac{(\vec{B} \cdot \nabla) \vec{B} } {|B^2|}.
 \label{eq:curv}
\end{equation}
When the magnetic field gets amplified by adiabatic compression, the curvature should stay almost constant because $B \propto S \propto 1/K$, where $S$ is a surface area. On the other hand, when a dynamo operates, the curvature will anti-correlate with the magnetic field strength. The evolution of curvature in a turbulent magnetic field can be predicted based on the Kasantsev theory \citep[e.g.][for a recent review]{2015PhRvE..92b3010S}, as in \citet{2002ApJ...576..806S}, who predicted stationary distribution of $K$ with a power-law slope $\propto K^{-13/7}$. 

\begin{figure}
\begin{center}
\includegraphics[width = 0.495\textwidth]{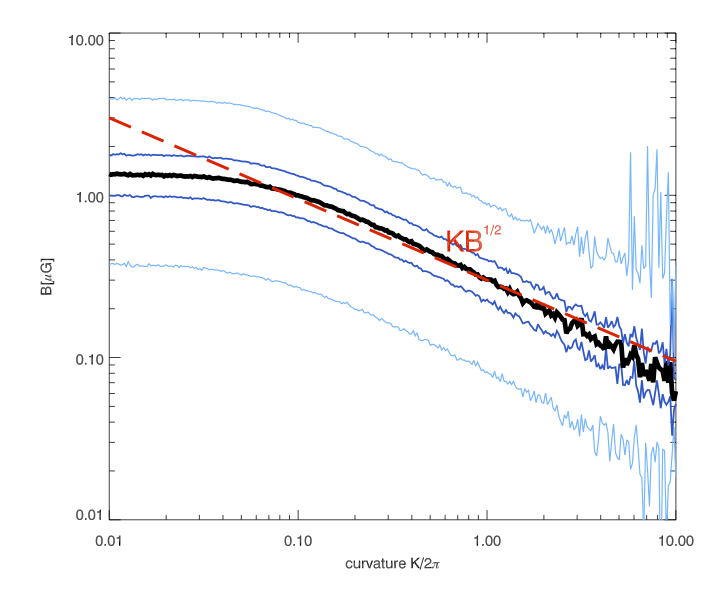}
\includegraphics[width = 0.495\textwidth]{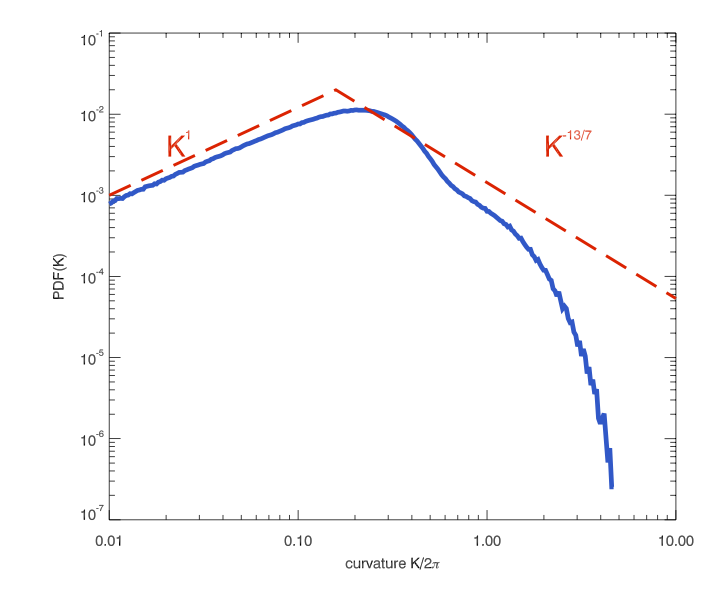}
\caption{Mean magnetic field as a function of curvature (top) and distribution function of curvature (bottom) for the innermost volume of our AMR8 run at $z=0$. In the top panel we show in dark and light blue the 1 and 2 $\sigma$ deviation around the mean (black).}
 \label{curv}
\end{center}
\end{figure}

In Fig.~\ref{curv}, we show the distribution of the curvature, $K$, in our AMR8 run, compared to the prediction by \citet[][]{2004ApJ...612..276S}: the  average curvature as a function of magnetic field (top) and the volume distribution function of $K$ in the innermost $\rm Mpc^3$ (bottom). Both  statistics show a good agreement with the results from small-scale dynamo theory across a wide range of scales. This supports the notion that amplified fields are indeed counter-acting the further bending of field lines with their  increased tension, consistent with the small-scale dynamo model.\\

Finally, we wish to investigate whether the final magnetic field observed in the AMR8 run at $z=0$ is energetically consistent with the kinetic turbulent budget available for the ICM. To this end, we repeated a similar
analysis as in  \citet{2016ApJ...817..127B}, i.e. we measured the kinetic energy flux across the turbulent cascade in our run, and compared its evolution to the measured growth of magnetic fields within the same volume.

\begin{figure}
\begin{center}
\includegraphics[width = 0.495\textwidth]{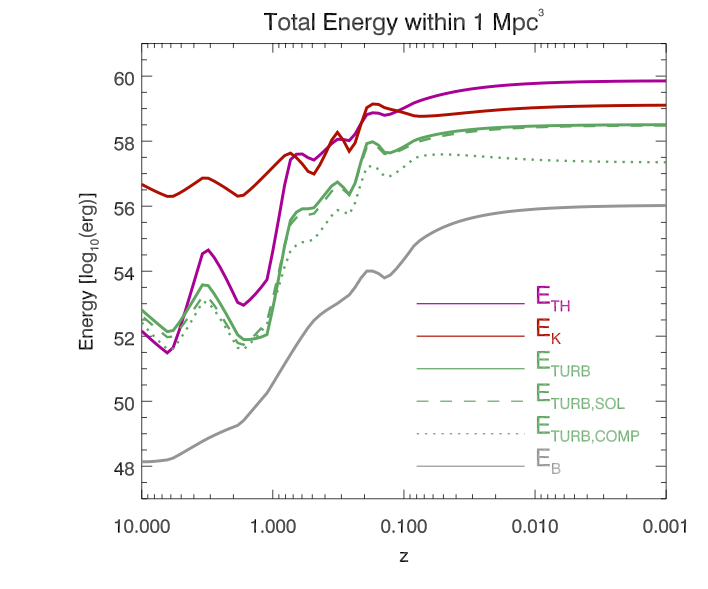}
\includegraphics[width = 0.495\textwidth]{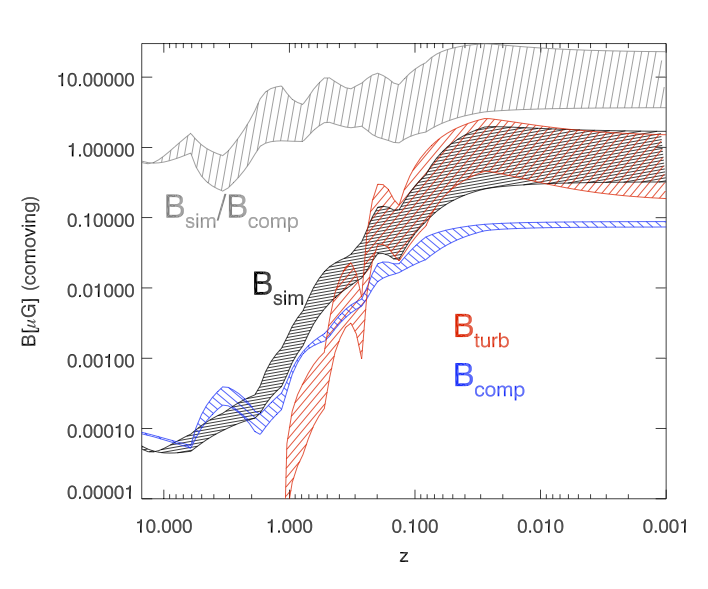}
\caption{Top panel: gas and magnetic field energy evolution within a $\rm Mpc^3$ comoving region around the cluster core in our AMR8 run. The total energy values shown here are for the thermal  energy ($E_{\rm TH}$), the  kinetic energy ($E_{\rm K}$), the small-scale filtered turbulent energy ($E_{\rm TURB}$, see Sec.~\ref{dynamo} for details), the solenoidal and compressive components of the turbulent energy ($E_{\rm TURB,SOL}$) and $E_{\rm TURB,COMP}$) and the magnetic field energy ($E_{\rm B}$).   Bottom panel: comparison between the magnetic field growth observed in our simulation and the theoretical expectations from a small-scale dynamo and gas compression. The black area gives the magnetic field strength in innermost comoving $\rm Mpc^3$, compared to the prediction from a frozen-in compression (blue) and from the dynamo amplification model by \citet{2016ApJ...817..127B}, assuming a $4 \%$  dynamo amplification efficiency. The dashed areas show the scatter of values obtained within the selected volume.}
\label{b_ev}
\end{center}
\end{figure}

The turbulent rms velocity is here measured by filtering out the large-scale velocity field, and the solenoidal component (relevant to the dynamo amplification) is extracted from the filtered field with a procedure similar to what outlined in \citet{va14mhd} and \citet{va17turbo}. In summary, we used a high-pass filter on the velocity field with 
a fixed $L = 100 ~\rm kpc$ scale {\footnote{In other works we proposed other filtering techniques to disentangle laminar from small-scale turbulent motions \citep[][]{va12filter,va17turbo}, yet for the purposes of this work this technique is accurate enough to measure the kinetic energy flux on small-scales in the central cluster regions. }}, and we FFT-transformed the turbulent velocity vector field, ${\vec V}(\vec k) = \mathcal{F}(\vec{v}(\vec r))$.
The solenoidal velocity component in Fourier space 
 ($\vec k\cdot{\vec v}_{\rm sol}(\vec k)=0$), is computed as  $\tilde{v}_{\rm i,sol}(\vec{k}) =\sum_{j=1}^3 \left( \delta_{i,j} - \frac{k_i k_j}{k^2} \right) \tilde{v}_j(\vec{k})$, and then the solenoidal component in real space is found via inverse FFT. 
The kinetic energy flux across the turbulent cascade measured in every cell is $\epsilon_s = 1/2 (\rho \sigma^3_{v,s} / L)$. 
In Fig.~\ref{b_ev} (top panel) we show the evolution of the total thermal, kinetic, turbulent (also in its compressive and solenoidal components) and of the magnetic energy within a comoving volume of $1 ~\rm Mpc^3$ centered on the cluster core.   The kinetic energy in the innermost cluster regions becomes stationary only in the last Gyr. During its late evolution $z \leq 0.1$ the kinetic energy in the innermost region of the cluster is $\sim 20\%$ percent of the thermal energy, while the small-scale turbulent energy is $\sim 5\%$ few percent. Note the clear dominance of the solenoidal component. By $z=0$ the magnetic energy is a few percent of the small-scale turbulent energy, and $\sim 1-2 \cdot 10^{-4}$ of the thermal energy, which corresponds to a $\beta_{\rm pl} \sim 1000$ (where $\beta_{\rm pl}$ is the ratio between thermal and magnetic pressure).
Following \citet{2016ApJ...817..127B}, the saturated magnetic field produced by the small-scale dynamo is:

\begin{equation}
B_{\rm turb} = [8 \pi \int_t C_E \epsilon_s dt]^{0.5},
\label{eq:Bturb}
\end{equation}
where $C_E$ is a small number $\sim O(10^{-2})$, for which slightly different values have been found by different authors \citep[][]{2015PhRvE..92b3010S,2015ApJ...810...93P,2016ApJ...817..127B}. We iterated Eq.~(\ref{eq:Bturb}) at every timestep, using $t$ as the elapsed physical time between two timesteps. In what follows, we will present the evolution
of $B_{\rm turb}$ and we will assume a constant $C_E=0.04$.

In Fig.~\ref{b_ev} we show the evolution of the magnetic field strength directly measured in the simulation, compared with 
the $B_{\rm turb}$ estimated as above and the magnetic field which can be obtained via simple compression ($B_{\rm comp} = B_0 (n/<n>)^{2/3}$) based on the gas density in the simulation and starting from our initial seed field.

On average, the magnetisation in the innermost $\rm Mpc^3$ (coincident with the cluster center at $z=0$) grows by
four orders of magnitude from $z=30$ to $z=0$. Based on the observed trend, the initial growth down to $z \approx 1$ is mostly explained by gas compression, and after this point the turbulent amplification takes over, amplifying the
volume-weighted field $\sim 10$ times above the compressed field level. At several redshifts, we observe
a clear correlation between blips in the magnetic field growth and in $B_{\rm turb}$, which suggests
that indeed close to these epochs turbulence is transported at small scales and the simulated magnetic fields
grows accordingly.

Overall, this comparison confirms that the development of turbulence and magnetic field growth are
tightly correlated for $z \leq 1$, and that the observed field growth is consistent with
an amplification efficiency of the order of a few percent, i.e. in the range of what is found in other works \citep[e.g.][]{2015PhRvE..92b3010S,2016ApJ...817..127B}.

\subsection{Comparison with observed Faraday Rotation Measure}
\label{RM}

\begin{figure}
\begin{center}
\includegraphics[width = 0.495\textwidth]{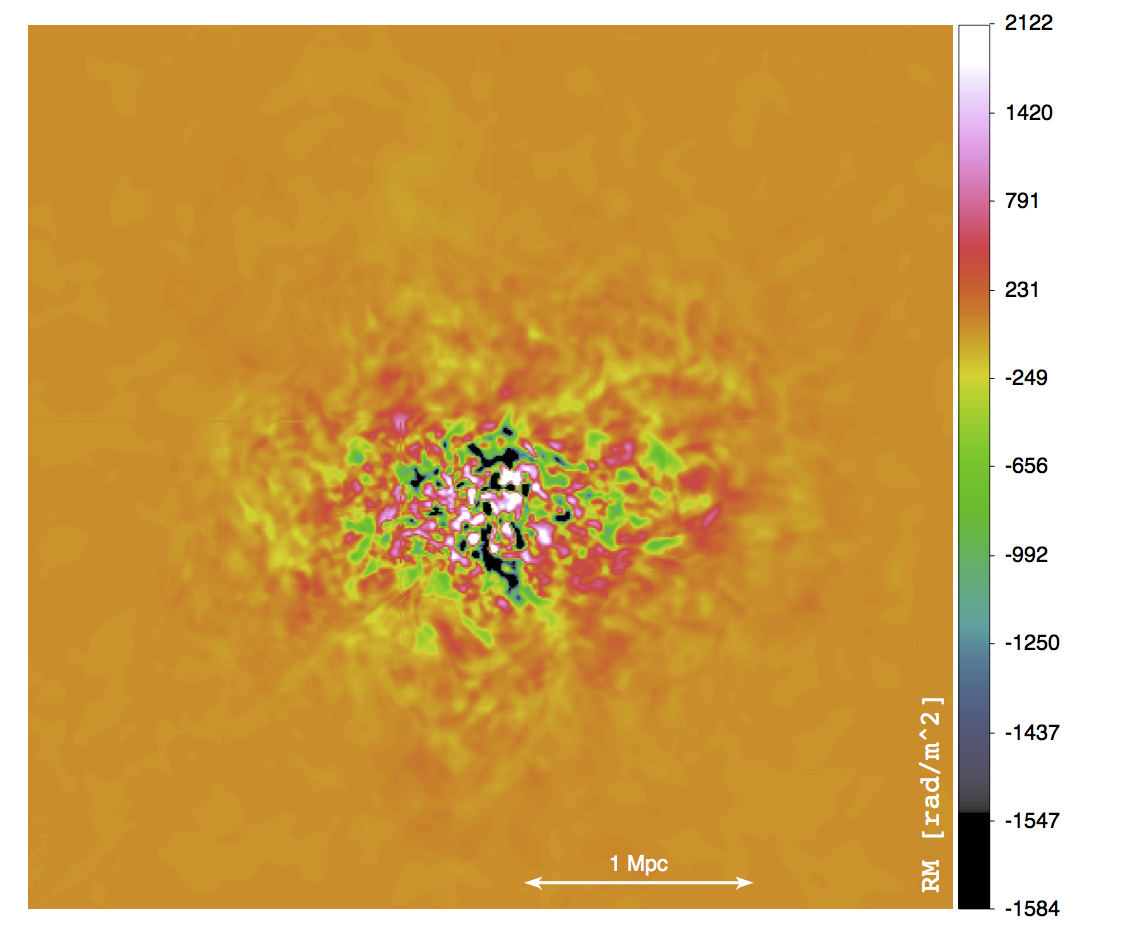}
\caption{Simulated map of Faraday Rotation from our cluster at $z=0$ (AMR8). }
 \label{RM_map}
\end{center}
\end{figure}

Finally, we produced projected maps of RM for our AMR8 run, and compared this 
model with the Very Large Array observations of the Coma cluster by \citet[][]{bo10} and \citet{bo13}. We computed the RM along each of the coordinate axes of the simulation, by measuring for each $(x,y)$ pixel:

\begin{equation}
  \rm RM(x,y)[rad/m^2]=812 \sum_l \frac{B_{\rm ||}(x,y,z)}{\rm \mu G} \cdot \frac{n(x,y,z)}{\rm cm^3} \Delta x,
\end{equation}
where $\rm ||$ denotes the component parallel to the line of sight and we then computed the mean of $\rm RM$ and the dispersion of $\rm RM$ within a reference area in order to compare with the observations available for the
Coma cluster \citep[][]{bo10,bo13}. An example of the RM map (along the same axis of the previous maps) is given in Fig.~\ref{RM_map}.
A detailed comparison with observations of the Coma cluster by \citet[][]{bo10} is not straightforward because of the different resolution of simulations (3.95 kpc) and observations ($\sim$ 0.7 kpc). Having the highest possible resolution is important to include the RM fluctuations produced by the magnetic fields on small spatial scales. 

In addition, the RM images analysed in \citet[][]{bo10} have a small size, and the high resolution of those observations was crucial to obtain RM values over several independent beams. \par
Degrading the resolution in RM studies has two potential disadvantages: (i) reducing the number of independent beams over which the RM can be derived. As a consequence, the sampling errors on $\langle RM \rangle$ and $\sigma_{RM}$ -- proportional to $\rm{(N_{beams}})^{-1/2}$ -- increase; (ii) increasing the beam depolarisation. Hence, the RM can be derived over less regions and with lower accuracy.
In the case of the Coma cluster, it is mainly (i) which prevents us from a detailed comparison with the simulations presented in this paper.
If observations are rescaled to the resolution of 3.95 kpc, no source has an adequate number of measurements of RM to derive $\langle RM \rangle$ and $\sigma_{RM}$ (i.e. a RM value in more than 3 independent beams).\par

\begin{figure*}
\begin{center}
\includegraphics[width = 0.495\textwidth]{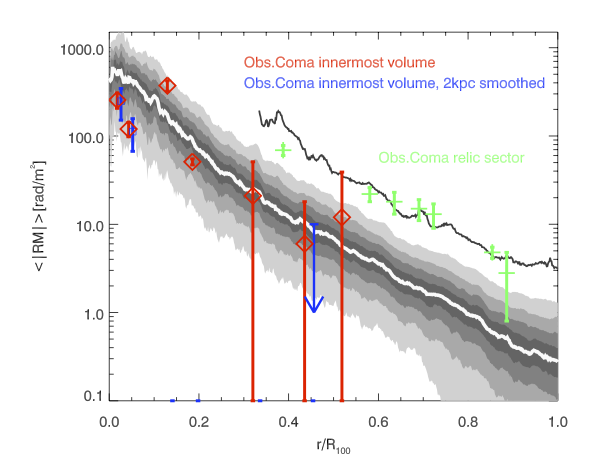}
\includegraphics[width = 0.495\textwidth]{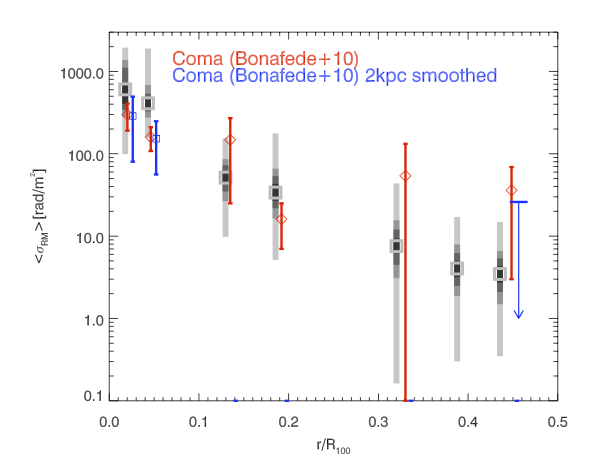}
\caption{Comparison between simulated and observed radial trends of Faraday Rotation. Left panel: radial profile of mean RM for our AMR8 run at $z=0$ {\bf(considering the distribution along three different coordinate axes)}. The shaded area shows the percentile distributions (from $10 \%$ to $90 \%$ while the colored points with errorbars show the observed data for the Coma cluster, considering the
  RM data for the innermost region \citet[][]{bo10} and limited to the outer arm where the Coma C source is \citet[][]{bo13}. The additional black line shows the mean RM profile in our simulation in a narrow sector running along a filamentary accretion within the cluster. 
  Right panel: profile of the dispersion of RM in our simulated mock RM map (same percentiles as in the previous panel) measured within areas equivalent to real observed sources in \citet{bo10}. The additional red points with errorbars show the original data from \citet{bo10}. In both panels we additionally show in blue the result of our re-gridding procedure at $2$ kpc of the original radio data from \citet{bo10}, see Sec.~\ref{RM} for more explanations.}
 \label{RM_prof}
\end{center}
\end{figure*}

The best compromise that we found is to compare the simulations with observations rescaled at the resolution of $\sim$ 2 kpc, which leaves enough independent measurements of RM per sources for 3 sources, and have a spatial resolution that is only a factor 2 (rather than 6) higher than the original data.\\
To derive the RM images at 2 kpc resolution, we started from the maps of Stokes Q and U and convolved them with a Gaussian beam having the FWHM of $4'' \times 4''$. Then, following the same procedure as explained in  \citet[][]{bo10}, we derived the RM images and the RM statistics.  The values of  $\langle RM \rangle$ and $\sigma_{\rm RM}$ are shown in Fig. \ref{RM_prof}. We note that the values of $\sigma_{\rm RM}$  are slightly smaller, and ~rrors are larger because of the increased sampling errors. For the source at $\sim 0.45 ~R_{\rm 100}$, our degrading procedure only yields upper limits in both quantities.

Fig.~\ref{RM_prof} shows the  two-dimensional profile of RM from the mock observation, considering the average profile of the three RM maps along the coordinate axes of the simulation.
Although all the caveats explained above should be kept in mind, the average profile of RM in the simulation is fairly similar to the observed trend of RM in the innermost Coma region, and all observed RM values are consistent with the simulated profile within the $20-80 \%$ confidence interval. Even in the outer region where \citet{bo13} probed the trend of RMs limited to a narrow sector located in the direction of the radio relic in the Coma cluster, our mean RM trend is compatible with observations. In this case, we computed the mean profile of RM along the narrow $\approx 0.5 ~\rm Mpc  \times 2 ~\rm Mpc$ filamentary accretion pattern north of our cluster (see Fig.~2), which gives the black line in the figure). The gas properties of this region of Coma are not very well constrained \citep[see discussion in][]{bo13}. Therefore, while a systematic comparison with simulated filaments is not yet possible, our simulated trend can only confirm that while the RM level in the relic sector of Coma is not compatible with the regular profile of the cluster, overdensities associated with filaments can indeed explain such large RM.\\

Others have predicted the distribution of RMs in large-scale structures, comparing simulated clusters to collections of observed RM from various clusters \citep[e.g.][]{donn09,2011ApJ...739...77X} or to a compilation of RM values from surveys in polarisation \citep[e.g.][]{2010MNRAS.408..684S,2017arXiv170601890S}. Comparing to a heterogeneous RM dataset is non-trivial because the effect of beam depolarisation affect objects at different distances to different degrees, and the scatter in RM of large datasets is increased by the presence of cool core and non-cool core system. 
Such problems are mitigated when comparing to a sample of RM from a single object, in which case the effect of beam depolarisation can be accounted for (however, the role of cosmic variance may be more significant).  To  our knowledge, no direct comparison to the observed RM profile of the Coma cluster (probably the object for which we have more information from continuum and polarisation radio data) has been performed before.  For example,  \citet{2011ApJ...739...77X} performed a similar  procedure to compare their simulated RM maps to observed ones. However, they compared to clusters at higher redshift and with data probing a lower frequency. As a result, the probed a smaller range of physical scales compared to our simulation.

In summary, the analysis of RM data from our best-resolved simulation confirms that the level of magnetic field amplification produced in our cosmological simulation is compatible with observations. Hence, it is conceivable that the magnetic
field in the Coma cluster comes from the dynamo amplification of primordial fields (in this case $0.1 ~\rm nG$). As far as we know, this is the closest a simulation has come to reproduce the RM profile of the Coma cluster starting from a realistic value for the primordial seed field.

\section{Discussion}
\label{sec:discussion}

\subsection{Numerical and physical uncertainties}
\label{numerics}

In the simulations discussed in this paper we neglected all physical mechanisms other than gravity and (magneto)hydrodynamics, which are otherwise crucial to model galaxy formation (e.g. radiative gas cooling, chemical evolution, star formation and AGN). The complex physical interplay of these mechanisms is numerically difficult to handle, and only a few works showed evidence of rather successful prescriptions for galaxy formation in cosmological simulations \citep[e.g.][]{2014Natur.509..177V,2017Galax...5...35D,2017MNRAS.470..166H,2017arXiv170206148H}. Moreover, the process of galaxy formation itself can contribute to the seeding of magnetic fields \citep[][]{2006MNRAS.370..319B,donn09, xu09,2017arXiv170601890S,va17cqg}. The purpose of focusing on non-radiative physics was to identify the properties of magnetic field amplification by a small-scale dynamo. In radiative simulations with feedback, it will be considerably more difficult to identify sources of magnetic field amplification. The fact that at a large enough spatial resolution even a non-radiative simulation can produce magnetic fields compatible with RM observations may suggest that large-scale magnetic fields in the intracluster medium can be explained purely by structure formation.  However, in future work we also plan to increase the realism of the simulations by gradually including more physical processes, as in \citet{va17cqg}.

\subsubsection{The role of spatial resolution}
The resolution and accuracy of our numerical scheme are the  essential points in our study.
In all our runs we relied on the Dedner cleaning algorithm \citep[][]{ded02}, whose main limitation is  the reduction of dynamical range achieved for a given grid size, due to the intrinsic dissipation of the scheme. 
Compared to other MHD method such as Constrained Transport, the Dedner scheme is more affected by small-scale dissipation of magnetic fields,  due to the $\nabla \cdot \vec{B}$ cleaning waves it generates to keep the numerical divergence under control \citep[][]{kri11}. 
Nevertheless, several works in the literature have shown that the 
method is robust and accurate for most idealized tests in MHD \citep[e.g.][]{wa09,wang10,enzo14}.
Additional works comparing this scheme to others in more realistic astrophysical applications concluded that this method converges to the right solution in idealized tests, unlike other common cleaning or $\nabla\cdot \vec{B}$ preserving  techniques \citep[][]{2013MNRAS.428...13S,2016MNRAS.455...51H,2016MNRAS.461.1260T}.

\begin{figure}
  \begin{center}
    \includegraphics[width = 0.495\textwidth]{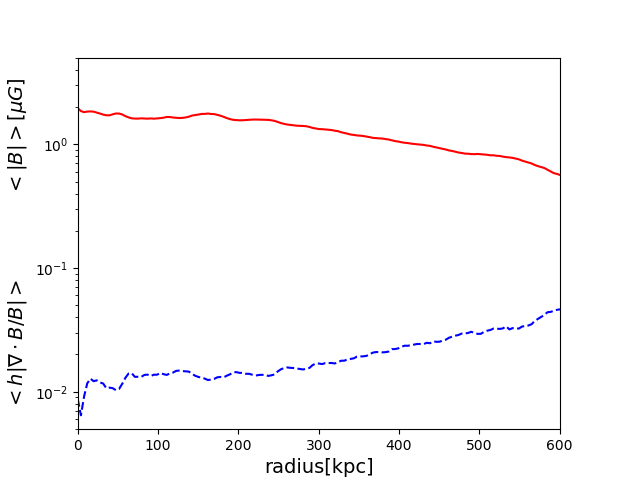}
    \caption{Tests on the degree of conservation of $\nabla \cdot \vec{B}$ of our MHD solver: radial volume-weighted profile of  $h|\nabla \cdot \vec{B}/B|$ for the AMR8 run (bottom), compared with the volume-weighted profile of magnetic field strength in the same volume (top).}
    \label{fig_div}
  \end{center}
\end{figure}

First, we verified in the AMR8 run at $z=0$ how well the $\nabla \cdot \vec{B}=0$ condition is preserved in our simulations. Fig.~\ref{fig_div} gives the radial (volume weighted) profile of $h|\nabla \cdot \vec{B}/B|$ for AMR8 run, where $h=2$ cells is the stencil used to compute the divergence with a simple first-order finite difference scheme,(i.e. $(\nabla \cdot \vec{B})_x=B_x(i+h/2)-B_x(i-h/2)$ for the x-component, etc). 
In the largest part of the simulation box, the numerical divergence is $\sim 2-3 \%$ of the local magnetic field value, which makes the level of spurious magnetic energy  $\leq 10^{-4}$ of the magnetic energy on larger scales. This confirms that in our application the numerical effects are small enough, and that the energy produced by the small-scale dynamo is much larger than local spurious fluctuations that may be caused by the Dedner scheme. \\

Estimating the typical kinematic and magnetic Reynolds numbers attained in these simulations is made non-trivial by many factors: gas flows are not stationary and characterised by several different scales, the system is not closed and the effective viscosity and resistivity are set by the numerical scheme, which has a variable spatial resolution because of AMR.  An upper limit on the Reynolds number in numerical flows can be obtained by assuming an ideal Kolmogorov model of turbulence \citep[e.g.][]{kri11}:

\begin{equation}
  R_{\rm e,max} \approx (0.5 L /\Delta x)^{4/3} ,
\end{equation}
where $L$ is the maximum correlation scale in the flow and $\Delta x$ is the spatial resolution. Limited to the central high-resolution region $L \approx$ 2 Mpc this estimate yields $R_{\rm e,max} \approx 1600$. Conversely, a lower
limit is given by assuming a first order numerical scheme  \citep[e.g.][]{2017arXiv170405845R}:

\begin{equation}
R_{\rm e,min} \approx L /\Delta x,
\end{equation}
yielding $R_{\rm e,min} \approx 500$ in this case. If we consider the entire virial volume ($L \approx 6 ~\rm Mpc$) at the resolution corresponding to the 6th refinement level (16 kpc), the above estimates yield $R_{\rm e,min} \approx 380$ and $R_{\rm e,max} \approx 1100$. 

Clearly, these estimates are still relatively crude as flows in the ICM are not stationary, turbulence and magnetic fields have intermittent distributions, and multiple injection scales can be present at the same time 
\citep[e.g.][]{va12filter,va17turbo}.

For the magnetic Reynolds number, we can in principle assume $R_{\rm e} \approx R_{\rm M}$, given that the artificial viscosity and resistivity are of the same order $P_M=R_{\rm M}/R_{\rm e}=\nu/\eta \approx 1$. However, the small-scale $\nabla \cdot \vec{B}$ waves generated by the Dedner scheme may reduce the {\it effective} $R_{\rm M}$ further. 
Given the rapid growth of the magnetic field, we conclude that at least towards the end of our most resolved runs the magnetic Reynolds number is large enough to allow for the development of a small-scale dynamo up to the final linear amplification stage. This implies that $R_{\rm M} \gg 100$. We notice that following similar considerations, \citet{2017arXiv170405845R} observed the onset of the linear growth stage of the small-scale dynamo inferring a minimum magnetic Reynolds number in the range $R_{\rm M} \sim 100-200$. 

Given that other MHD methods less prone to numerical dissipation can reach the same dynamical range at a $\sim 2-4$ coarser effective resolution, we suspect that similar magnetic Reynolds numbers can be achieved by higher-order MHD methods at a spatial resolution of $\sim 8-16$ kpc, provided that the initial seed fields are similar to what we have assumed.\\

\begin{figure}
  \begin{center}
    \includegraphics[width = 0.495\textwidth]{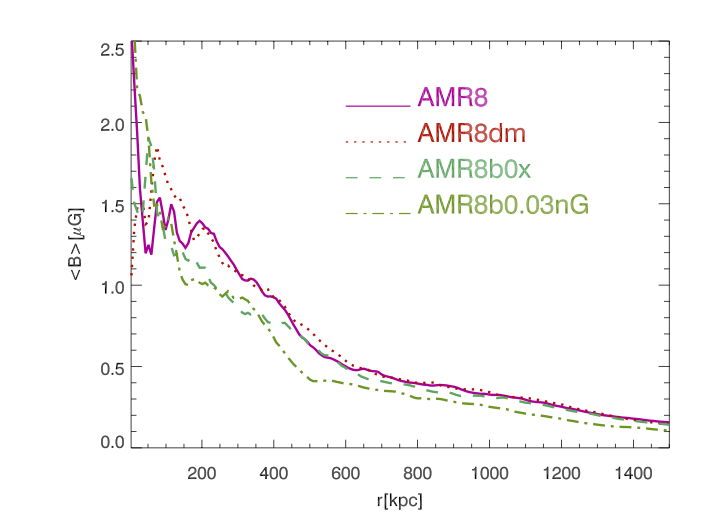}
    \caption{Radial profiles of the median magnetic fields at $z=0$ for additional resimulations testing the role of the initial seed field (AMR8b0x and AMR8b0.03nG, see Tab.~1 for details) and of the mass resolution (AMR8dm).}
    \label{fig_prof2}
  \end{center}
\end{figure}

\begin{figure}
  \begin{center}
    \includegraphics[width = 0.495\textwidth]{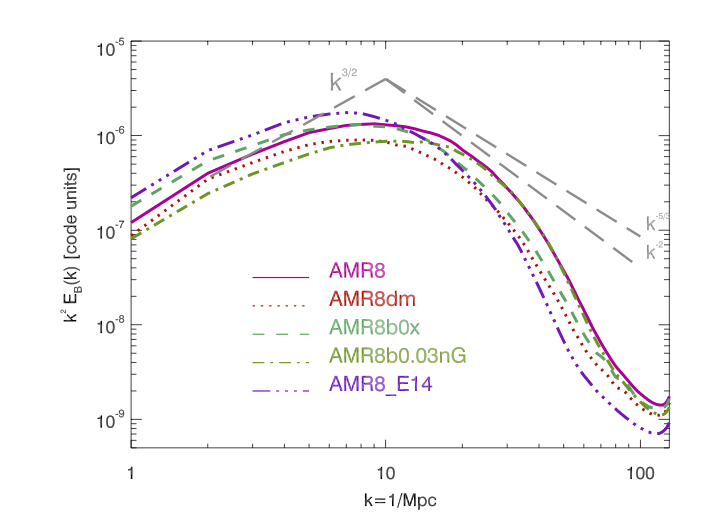}
    \caption{Spectra of magnetic field energy at $z=0$ for runs starting with an initial seed field (AMR8b0x and AMR8b0.03nG), with an increased mass resolution (AMR8dm), and for a relaxed galaxy cluster (AMR8dm\_E14, see Tab.~1 for details).}
    \label{fig_spectra2}
  \end{center}
\end{figure}

\subsubsection{The role of mass resolution}
Our limited mass resolution for dark matter particles ($m_{\rm DM}=1.04 \cdot 10^{11} M_{\odot}$) may  prevent the injection of turbulence from self-gravitating satellites due to the shallower potential which can form if the mass resolution is coarse. 
Self-gravitating gas substructures are an important driver of turbulence \citep[e.g.][]{in08,2017arXiv170602315W} and the resolution for DM particles we used here is significantly larger than in our previous work \citep[][]{va10kp, va14mhd}.
In order to study the effect of an increased mass resolution, we resimulated the AMR8 run with an eight times increased mass resolution (run AMR8dm), i.e. $m_{\rm DM}=1.3 \cdot 10^{10} M_{\odot}$, using two levels of nested initial conditions as in \citet{va10kp}, which introduce twice as many DM particles. We find that the resulting three-dimensional distribution of magnetic fields at $z=0$ does not show significant differences compared to our reference AMR8 run, at least the modest scatter which is expected for the increased number of substructures in the run with more DM resolution: the radial profile of magnetic field is consistent with the reference AMR8 run at all radii (Fig.~\ref{fig_prof2}), and also the power spectrum of the magnetic energy has nearly the same shape and normalisation (Fig.\ref{fig_spectra2}). We conclude that, while the increase in DM resolution is surely mandatory to properly resolve galaxy formation within simulated clusters, its impact on the simulated magnetic fields is minor. \\

\subsubsection{Sensitivity to the initial seed field}
In order to assess the dependence of our final field configuration on the amplitude and geometry of the initial seed field we performed a few more resimulations of the AMR8dm case: in particular we tested a lower initial field of $0.03~\rm  nG$ at $z=30$ (AMR8b0.03nG) or an initial field of $0.1 ~\rm nG$ as in AMR8dm but only aligned in the X-direction (AMR8bx). 
As can be seen in Fig.\ref{fig_prof2} and Fig.\ref{fig_spectra2}, the radial field distribution and the spectral properties of the fields at $z=0$ are very similar to the AMR8dm case, suggesting at the highest resolution our simulated cluster is in the non-linear dynamo regime indeed, and that the field configuration in the innermost cluster regions is fairly independent on the seed field \citep[e.g.][]{2015MNRAS.453.3999M,2016MNRAS.456L..69M}. This is consistent with the idea that the origin of cosmic magnetism is better investigated in cluster outskirts or in filaments \citep[][]{va17cqg}. We shall note, however, that for initial field strengths $\leq 0.03 ~\rm nG$ the final field in our cluster gets increasingly smaller, suggesting that for seed fields below this threshold our resolution is not enough to properly resolve the $l_{\rm A}$ for most of the cluster evolution, and the  non-linear amplification regime develops too late (or never begins), as noted by \citet[][]{2016ApJ...817..127B}. Only with future (even more resolved) simulations we will be able to test to which extend is the final field configuration in simulated clusters independent on the assumed primordial seed field.

\subsubsection{Comparison with a relaxed cluster}
\begin{figure}
  \begin{center}
    \includegraphics[width = 0.495\textwidth]{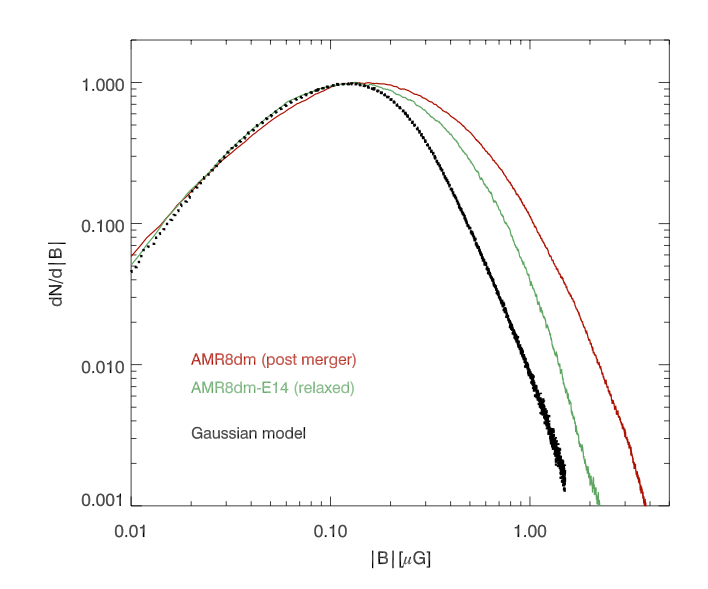}
    \caption{Distribution of magnetic field strength in objects with a different dynamical state: we show the PDF of the magnetic field strength for the central $(1.25 ~\rm Mpc)^3$ region of our AMR8dm run (the perturbed cluster studied in the main paper) and for the AMR8dm\_E14 simulation of a relaxed cluster with a similar mass, in both cases at $z=0$. The Gaussian distribution matching the peak of the two simulated PDF is additionally given as a black line.}
    \label{fig_PDF2}
  \end{center}
\end{figure}

Finally, we present a second cluster with a similar final mass, simulated with an identical AMR scheme and a DM mass resolution as in the AMR8dm (run AMR8dm\_E14 in Tab.~1). This cluster has a total virial mass of $\approx 1.0 \cdot 10^{15} M_{\odot}$ at $z=0$, but has a very different dynamical history compared to our fiducial cluster. In particular, this cluster shows no evidence of a major merger for $z \leq 1$ and is the most relaxed system in the sample \cite[][]{va10kp}. 
The magnetic power spectrum has a similar shape as the reference AMR8dm case (Fig.\ref{fig_spectra2}), but with a slightly larger maximum scale for the field ($\sim 200$ kpc compared to $\sim 100$ kpc), possibly suggesting an earlier start of the dynamo amplification. Again, the field strengths do not follow a Maxwellian distribution (Fig.~\ref{fig_PDF2}), even if the tail of the distribution is less pronounced than in the AMR8dm run. This is consistent with the view that the specific merger history of each cluster determine the amount of multiple ($\sim$Gaussian) components which co-exist in the intracluster medium at a given time. although an extensive study of magnetic field configurations for clusters with different dynamical states and masses is deferred to future work.

\subsection{Observational perspectives}

Whereas turbulence-in-a-box simulations of a dynamo have highlighted the potential of RM to constrain the micro-physics of the ICM \citep[e.g.][]{2016MNRAS.455.3702N,2017MNRAS.465.4866S}, our work suggests the presence of departures from the commonly assumed Gaussian model of ICM magnetic fields, even on large scales. 
Even though the number of RMs that are typically observed per clusters is still rather small (maximum $7$ sources at $\sim \rm Mpc$ distance from the center of Coma, and $7$ in the SW sector of the cluster), future radio surveys can test our results \citep[e.g.][]{2015aska.confE..95B,2015aska.confE..92J}.

\begin{figure*}
  \begin{center}
    \includegraphics[width = 0.95\textwidth]{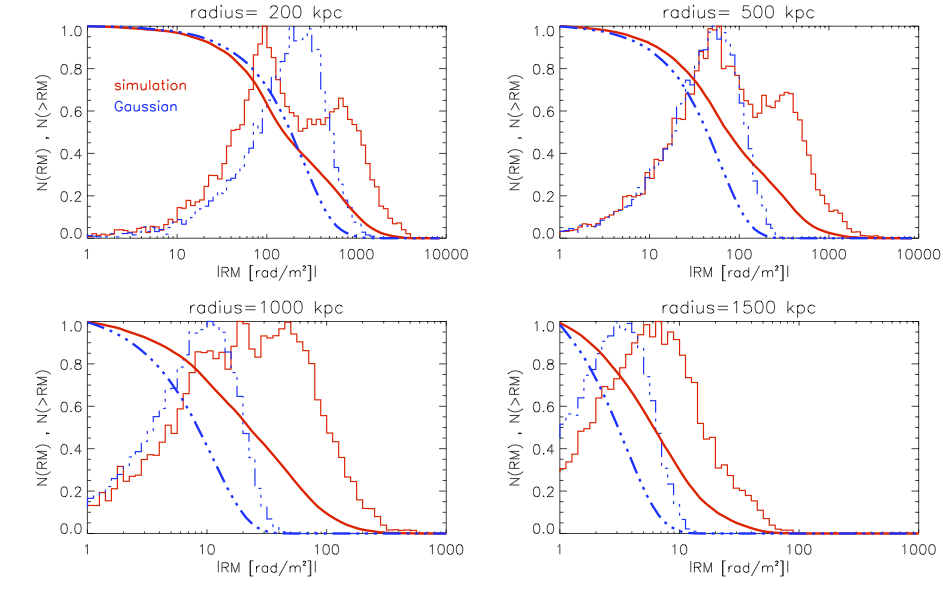}
    \caption{Analysis on the possibility of observing departures from Gaussianity in the observed distribution of RM: the histograms show the differential distribution function of RMs at several distances from the cluster center of our AMR8 run (red lines) compared to the differential  distribution expected from a Gaussian model  as in \citet{bo10} . Each distribution is computed considering a shell of $200 ~\rm kpc$ at each radius. The thick lines show the cumulative distribution function of RMs within the same circular shells.}
    \label{fig:pdf_rm}
  \end{center}
\end{figure*}

In Fig.~\ref{fig:pdf_rm}, we show how the distribution of RMs (considering only the projection along one axis) compares between our AMR8 model and the Gaussian case at four different distances from the cluster centre (assuming projected shells of 200 kpc width). While the presence of non-Gaussian features in the distribution of RMs is present at all radii, the difference relative to the Gaussian model becomes more significant at radii $\geq 1 \rm ~Mpc$   ($\geq 0.3 R_{\rm 100} $ ), where the peak of the RM distribution in the AMR8 model clearly differs from the Gaussian scenario. At radii $\geq 1 \rm ~Mpc$, we predict no RM larger than $\sim 40-50 ~\rm  rad/m^2$ in the Gaussian mdoel, while a non-negligible fraction of background sources ($\sim 20-30 \%$) should have larger values if the magnetic field is non-Gaussian.  At radii $\geq 0.5 ~\rm R_{\rm 100}$ , we should have a few detections of RM $\geq 10 \rm ~rad/m^2$ only in the non-Gaussian case. 

In order to produce the necessary statistics from observations (i.e. a few tens of bright polarised background sources in the outer region of nearby clusters), we will have to 
wait for the next generation of radio telescopes as present facilities (e.g. the Jansky Very Large Array) require very long integration times to reach a $3 \sigma$ sensitivity higher than $\approx 10 ~\rm rad/m^2$. Moreover, very few polarised sources are expected per square degree, making the creation of a finely spaced RM grid -- even around local clusters -- extremely difficult \citep{Rudnick14}. Finally, only clusters at high and low Galactic latitudes are suitable targets, as the Galactic RM can easily hide a difference in the RM of the order of 10 $\rm ~rad/m^2$. \par
On the other hand, the SKA-MID should be able to recover the radial dependence of RMs of polarised sources behind clusters up to a large radius, far beyond the capabilities of current instruments \citep[e.g.][]{2013A&A...554A.102G,2015arXiv150102298T,2015aska.confE.105G,2015aska.confE..92J,2015aska.confE..95B}. 
In particular, the planned deep polarization survey with SKA-MID 
 is expected to detect between $\sim 300$ and $\sim 1000$ polarised
sources per square degree at 1.4 GHz \citep[e.g.][]{2015aska.confE..95B,2015arXiv150102298T}. 
For a Coma-sized galaxy cluster, this corresponds to measuring RM on $\sim 50$ background sources, with a formal error of the order of few $\rm rad/m^2$. \par
Before the advent of SKA1, deep surveys in polarisation with ASKAP (Possum) and Meerkat (Mightee-Pol survey) may approach this limit. 

Even in the presence of non-Gaussian magnetic fluctuations, as predicted by these simulations, the detection of RMs from sources in the cluster outskirts will depend on the amplitude of the seed field.
In these external regions, we expect dynamo amplification to be small \citep[e.g.][]{ry08,donn09,va14mhd}, the detection of significant RMs can be used to rule out competing scenarios.  For example, 
 the systematic detection of RMs $\geq 10 ~\rm rad/m^2$ at the virial radius of  galaxy clusters and separated from local sources of magnetisation such as radio galaxies can only be explained by the presence of primordial seed fields of the order of $0.1-1~ \rm nG$, or by an anomalous amplification of magnetic fields beyond what can be currently resolved by simulations. At  present, such large values of RM have been measured only along the SW sector of the Coma cluster \citep[][]{bo13}, which makes it difficult to derive general conclusions. \\

\subsection{Comparison with previous MHD simulations}

The results shown in this paper are in line, both, with our previous work on the subject \citep[][]{va14mhd} as well as with earlier non-radiative AMR MHD simulations \citep{br05,2008A&A...482L..13D,co11}. The latter demonstrated the growth of magnetic fields beyond what is be achieved by compression, albeit without clear evidence of having reached the non-linear amplification stage, consistent with the fact that the dynamical range achieved in our newest runs is larger. 
As a result of over-cooling, runs including radiative losses reported larger  magnetic fields ($\sim ~\rm \mu G$) in clusters starting from similar primordial seed field
\citep[][]{2008A&A...482L..13D,co11,ruszkowski11}. \citet[][]{xu09} and \cite{2011ApJ...739...77X} also reported indications of dynamo amplification in MHD simulated clusters, but it is difficult to relate their results to ours, due to the entirely different seeding mechanism (e.g. strong seeding by AGN at low redshift). 

A few recent papers, investigated the amplification of primordial magnetic fields in large-scale structure, using either a moving stencil method \citep[][]{2015MNRAS.453.3999M}  or a mesh-less Lagrangian technique \citep[][]{2016MNRAS.455...51H} in the cosmological context. The MHD methods applied in both cases are qualitatively similar to our choice here (i.e. divergence "preserving" or "cleaning" methods) and in both cases large values of magnetic field amplification in clusters are reported, even though it is difficult to say which stage of the dynamo amplification regime is attained in the two cases. 

Finally, we find some disagreement with the results reported by cosmological SPH
simulations \citep[][]{do99,do08,donn09,2009MNRAS.398.1678D,bo11,beck12,beck13}. There, already at high redshifts ($z \geq 2$), larger amplification factors for the magnetic energy are found, and this despite the seemingly smaller Reynolds number achieved in these simulations. Understanding these differences is beyond our goal here, and we can only speculate that the effective Reynolds in SPH simulations may not be entirely understood.

\section{Conclusions}

We have presented evidence for {\it resolved} dynamo growth of intracluster magnetic fields in cosmological grid simulations. This was obtained with high-resolution re-simulations of a Coma-like galaxy cluster, using the MHD version of {\enzo} \citep[][]{enzo14} and  aggressive adaptive mesh refinement. \\

Our simulations covered an unprecedented dynamical range in the innermost region of a cluster and showed evidence of a small-scale dynamo and local amplification of magnetic fields up to values similar to what is found in observations. Starting with a weak seed field  of 0.1 nG (comoving) at $z=30$, in cluster centres the magnetic fields approach energy equipartition with the kinetic energy flow on $\leq$ 100 kpc scales, and display clear spectral signatures consistent with the standard small-scale dynamo theory \citep[e.g.][]{2015PhRvE..92b3010S,2002ApJ...576..806S}. 
At our best resolution (3.95 kpc), we constrain an overall efficiency of order $\sim 4\%$ in the transfer between turbulent kinetic energy (in the solenoidal component) and the magnetic energy field. Our best run reaches a typical magnetic field level of $\sim 2 ~\rm \mu$G in the innermost Mpc$^3$ region, starting from an initial magnetic field of 0.1 nG (comoving), with maxima of $\sim 10 ~\mu$G. 

In flows with an effective Reynolds number much larger than what we achieved here (e.g.  \citealt{bl11b,2014ApJ...781...84S}), the  dynamical timescale to go from the kinematic to the non-linear growth regime is greatly reduced, from $\sim \rm Gyr$ to $\sim \rm kyr$ in case the Reynolds number is  $R_{\rm e} \sim 10^{12}$ \citep[][]{2016ApJ...817..127B}.  Hence, the  non-linear amplification that we see at low redshifts might have started much earlier, and the final field might be stronger and have larger spatial scales than what we found.  Yet even within the present limitations mirrored by the Reynolds number, the efficiency of the transfer of kinetic energy into magnetic energy in the innermost cluster regions at $z \sim 0$ is  $\approx 4\%$ and thus close to the one derived by \citet{2016ApJ...817..127B} for the saturated stage of dynamo amplification. 

Moreover,  the topology of the magnetic fields seem to be consistent with the most stringent observational constraints for the Coma cluster. In particular,  the Faraday Rotation of background polarised sources is in good agreement with the observations of RMs from real sources located behind Coma. This applied to, both, the average RM profile and its dispersion, even if in the latter case the comparison is limited to the first four sources owing to resolution effects.\\

A significant result is the fact that the RM observations appear to be reproduced by a significant non-Gaussian distribution of magnetic fields. Interestingly, these magnetic fields show a radial profile with a $\sim 3$ times lower normalisation
than what is usually inferred from these observations \citep[][]{bo10,bo13}. 
The departures from Gaussianity get more significant with increasing resolution, while the opposite trend is usually found in more idealised turbulence-in-a-box simulations \citep[][]{2013MNRAS.429.2469B,2014ApJ...781...84S}. 
Our results are explained by the superposition of different magnetic field components along the line-of-sight, which in turn makes the inversion of any observed RM trend into a three-dimensional model of magnetic field more complicated than in a single Gaussian component model.  

Here we focused on the amplification of one value for the initial magnetic field
and simulated only the dynamo amplification caused by turbulence induced by structure formation. Therefore, additional sources of turbulence and of magnetisation, such as active galactic nuclei, galactic winds and shocks, have been neglected. While the injection of turbulence on cluster-wide scales is still expected to be dominated by structure formation processes \citep[e.g.][]{su06,va12filter}, it is not clear what role other sources of magnetisation play \citep[e.g.][and references therein]{wi11,ry11,2013MNRAS.436..294B,2015ApJ...808...65F}.
The application of high-resolution MHD simulation to the study of extragalactic magnetic fields and to the prediction of their observational signatures will be essential to interpret future radio observations that aim to reveal the origin of cosmic magnetism.

\section{Acknowledgements}

The cosmological simulations described in this work were performed using the {\enzo} code (http://enzo-project.org), which is the product of a collaborative effort of scientists at many universities and national laboratories. We gratefully acknowledge the {\enzo} development group for providing extremely helpful and well-maintained on-line documentation and tutorials.\\
F.V. acknowledges financial support from the European Union's Horizon 2020 research and innovation programme under the Marie-Sklodowska-Curie grant agreement no.664931 and under the ERC Starting Grant "MAGCOW", no. 714196.
G.B.  acknowledges partial support from PRIN INAF 2014.  A.B. acknowledges support from the ERC Starting Grant "DRANOEL", no. 714245. 
We acknowledge the  usage of computational resources on the JURECA cluster at the at the Juelich Supercomputing Centre (JSC), under projects no. 11823, 10755 and 9016, and on the Piz-Daint supercluster at CSCS-ETHZ (Lugano, Switzerland) under project s701 and s805.  Finally, we wish to thank A. Shukurov and A. Seta for fruitful scientific discussions, and our anonymous referee for the helpful comments and suggestions that greatly improved the quality of our manuscript.

\bibliographystyle{mnras}
\bibliography{franco}

\end{document}